\documentclass[twocolumn,abs,prb,showpacs]{revtex4-1}

\usepackage{graphicx}
\usepackage{dcolumn}
\usepackage{float}
\usepackage{amsmath}
\usepackage{bm}
\usepackage[mathscr]{euscript}

\newcommand{\comment}[1]{}

\begin{document}


\title{Magnetic Properties of Dirac Fermions in a Buckled Honeycomb Lattice}

\author{C. J. Tabert$^{1,2,3}$}
\author{J. P. Carbotte$^{4,5}$}
\author{E. J. Nicol$^{1,2,3}$}
\affiliation{$^1$Department of Physics, University of Guelph,
Guelph, Ontario N1G 2W1 Canada} 
\affiliation{$^2$Guelph-Waterloo Physics Institute, University of Guelph, Guelph, Ontario N1G 2W1 Canada}
\affiliation{$^3$Kavli Institute for Theoretical Physics, University of California, Santa Barbara, CA 93106 USA}
\affiliation{$^4$Department of Physics, McMaster University,
Hamilton, Ontario L8S 4M1 Canada} 
\affiliation{$^5$Canadian Institute for Advanced Research, Toronto, Ontario M5G 1Z8 Canada}
\date{\today}

\begin{abstract}
{We calculate the magnetic response of a buckled honeycomb lattice with intrinsic spin-orbit coupling (such as silicene) which supports valley-spin polarized energy bands when subjected to a perpendicular electric field $E_z$.  By changing the magnitude of the external  electric field, the size of the two band gaps involved can be tuned, and a transition from a topological insulator (TI) to a trivial band insulator (BI) is induced as one of the gaps becomes zero, and the system enters a valley-spin polarized metallic state (VSPM).  In an external magnetic field ($B$), a distinct signature of the transition is seen in the derivative of the magnetization with respect to chemical potential ($\mu$) which gives the quantization of the Hall plateaus through the Streda relation.  When plotted as a function of the external electric field, the magnetization has an abrupt change in slope at its minimum which signals the VSPM state.  The magnetic susceptibility ($\chi$) shows jumps as a function of $\mu$ when a band gap is crossed which provides a measure of the gaps' variation as a function of external electric field.  Alternatively, at fixed $\mu$, the susceptibility displays an increasingly large diamagnetic response as the electric field approaches the critical value of the VSPM phase.  In the VSPM state, magnetic oscillations exist for any value of chemical potential while for the TI, and BI state, $\mu$ must be larger than the minimum gap in the system.  When $\mu$ is larger than both gaps, there are two fundamental cyclotron frequencies (which can also be tuned by $E_z$) involved in the de-Haas van-Alphen oscillations which are close in magnitude. This causes a prominent beating pattern to  emerge. 
}
\end{abstract}


\maketitle

\section{Introduction}

With the successful isolation of graphene in 2004, two-dimensional (2D) systems began to attract considerable attention.  While graphene provides a platform for investigating the physics of massless relativistic fermions\cite{Geim:2007,Neto:2009,Abergel:2010,Goerbig:2011,dasSarma:2011}, other 2D crystals are increasing in popularity as they promise to exhibit exciting phenomenon beyond those found in the single layer of graphite.  One of the more prominent extensions of graphene research has been the 2D topological insulators (TIs) introduced in the seminal paper by Kane and Mele\cite{Kane:2005}.  In a 2D TI, the nontrivial topology of the band structure allows helical edge channels to exist at the boundary of the sample.  These channels are topologically protected, and are robust against time-reversal-symmetry preserving impurities.  Recently, Hg$_x$Cd$_{1-x}$Te quantum wells have shown signatures of the helical edge-conduction\cite{Konig:2007,Roth:2009}.   

Another promising class of 2D TIs is the low-buckled honeycomb lattice with intrinsic spin-orbit coupling.  At low energy, these systems are expected to map onto the Kane-Mele Hamiltonian for the quantum spin-Hall insulator (QSHI)\cite{Kane:2005}.  Two candidate materials are silicene (with a spin-orbit band gap of $\Delta_{\rm so}\approx 1.55-7.9$ meV\cite{Liu:2011, Liu:2011a, Drummond:2012}), and germanene ($\Delta_{\rm so}\approx 24-93$ meV\cite{Liu:2011,Liu:2011a}).  The low-buckling of the honeycomb lattice causes the A, and B sublattices to sit in vertical planes with a separation distance $d$ ($\approx0.46$\AA\, for silicene\cite{Drummond:2012,Ni:2012}). In the presence of an external electric field oriented perpendicular to the system, the A-B asymmetry causes a potential difference $\Delta_z=E_z d$ to arise between the sublattices.  This spin splits the energy bands, and allows the resulting band gaps to be tuned.  It has been argued\cite{Drummond:2012,Ezawa:2012a} that, as $\Delta_z$ becomes greater than $\Delta_{\rm so}$, the system transitions from a TI to a trivial band insulator (BI).  At the critical value $\Delta_z=\Delta_{\rm so}$, the lowest band gap closes into a Dirac point, and the system is referred to as a valley-spin polarized metal (VSPM)\cite{Ezawa:2012}.  For $\Delta_{\rm so}$=8 meV, the critical electric field associated with the VSPM is $E_z\approx 17.4$ meV/\AA\, (a value often quoted in the literature\cite{Drummond:2012,Ezawa:2012,Ezawa:2012a, Ezawa:2012c}).  Recent theoretical studies have examined the effect of varying $\Delta_z$ on the AC conductivity\cite{Ezawa:2012b, Stille:2012}, magneto-optical conductivity\cite{Tabert:2013a,Tabert:2013c}, spin, and valley Hall effects\cite{Dyrdal:2012,Tahir:2013a,Tabert:2013b}, polarization function\cite{Tabert:2014,Chang:2014,VanDuppen:2014}, anomalous spin Nernst effect\cite{Gusynin:2014}, and quantum oscillations\cite{Islam:2014,Tsaran:2014,Shakouri:2014}.

The magnetization of Dirac-like materials has long been known to be anomalous\cite{McClure:1956,Fukuyama:1971,Safran:1979,Sharapov:2004,Gomez:2011,Principi:2010}.  A recent study\cite{Raoux:2014} showed how the orbital susceptibility of massless Dirac fermions on a 2D $\mathcal{T}_3$ lattice evolves from diamagnetic to paramagnetic as a function of the coupling strength between the honeycomb lattice, and an additional carbon atom placed at the center of each hexagon (i.e. going from the graphene lattice to the dice lattice).  For undoped graphene, the susceptibility shows a divergence\cite{Sharapov:2004,Gomez:2011,Principi:2010,Raoux:2014,Koshino:2011} as a function of the inverse magnetic field.  This is lifted by finite temperature\cite{Principi:2010} or other complications; these include the formation of a gap\cite{Sharapov:2004,Koshino:2011} which leads to massive Dirac fermions, effects of disorder\cite{Nakamura:2007}, or by finite doping away from the charge neutral point\cite{Sharapov:2004,Nakamura:2007}.  In the low-buckled honeycomb lattice discussed herein, we work with massive Dirac fermions except for the special case of the VSPM for which one of the gaps closes.  In this paper, we study the magnetic properties of such a system\cite{Ezawa:2012d} with particular emphasis on the changes in magnetization as one goes from the nontrivial to trivial topological phase.

Our paper is organized as follows: In Sec.~II, we present the theoretical background for the low-energy model of a buckled honeycomb lattice with intrinsic spin-orbit coupling.  Section~III contains a discussion of the grand thermodynamic potential, and results for the magnetization as a function of $\mu$.  We compute the $\mu$ derivative of the magnetization which is related to the quantized Hall conductivity through the Streda formula.  We also explore the plateaus which emerge in the integrated density of states, and optical spectral weight for finite magnetic field.  In Sec.~IV, we extend the results of gapped graphene to our system, and examine the magnetization, and magnetic susceptibility.  Finite temperature, and impurity effects are also considered.  Section~V contains our results for the magnetic oscillations at low field values.  Our conclusions follow in Sec.~VI.  

\section{Low-Energy Band Structure}

\begin{figure}[h!]
\begin{center}
\includegraphics[width=1.0\linewidth]{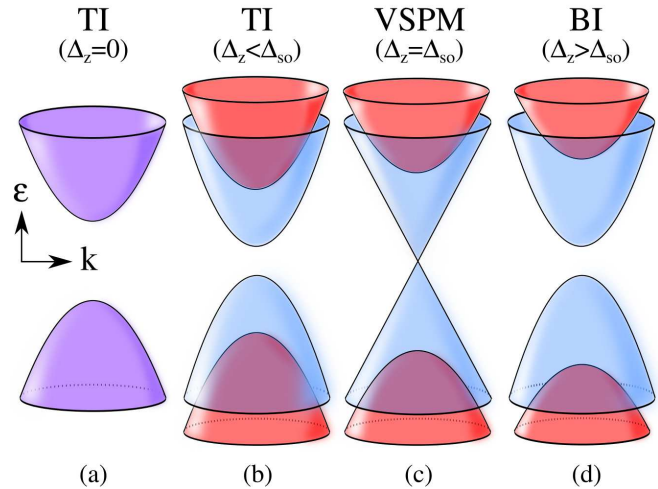}
\end{center}
\caption{\label{fig:Energy}(Color online) Schematic representation of the low-energy band structure of silicene about the $K$ point for varying electric field strength.  The purple bands represent spin degeneracy; the blue bands correspond to spin-up; and the red bands describe spin-down. Finite $\Delta_z$ spin splits the energy bands.  For $\Delta_z<\Delta_{\rm so}$, the system is a TI.  For $\Delta_z>\Delta_{\rm so}$, the system is a trivial BI.  At the critical value $\Delta_z=\Delta_{\rm so}$, the lowest gap closes, and the system is a VSPM. The spin labels are reversed at the $K^\prime$ point. 
}
\end{figure}

In the presence of an external electric field, the physics of a low-buckled honeycomb lattice with intrinsic spin-orbit coupling is well described by a simple nearest-neighbour-tight-binding Hamiltonian\cite{Liu:2011a, Ezawa:2012, Ezawa:2012a, Ezawa:2012b}.  At low energy, an effective Hamiltonian can be written\cite{Drummond:2012} to describe the physics in the vicinity of the two $K$ points of the hexagonal first Brillouin zone.  This Hamiltonian is of the well known Kane-Mele type\cite{Kane:2005} for describing the QSHI.   Written for a single spin, and valley,
\begin{align}\label{Ham}
\hat{H}_{\xi\sigma}=\hbar v(\xi k_{x}\hat{\tau}_{x}+k_{y}\hat{\tau}_{y})-\xi\sigma\frac{1}{2}\Delta_{\rm so}\hat{\tau}_{z}+\frac{1}{2}\Delta_{z}\hat{\tau}_{z},
\end{align}
where $\hat{\tau}_i$ are Pauli matrices associated with the pseudospin of the system.  The two valleys $K$, and $K^\prime$ are indexed by $\xi=\pm 1$, respectively, with $\hbar k_x$, and $\hbar k_y$ being the momentum components measured relative to the $K$ points.  The real spin of the electrons is given by $\sigma=\pm 1$ for spin-up, and -down, respectively.  The first term of Eqn.~\eqref{Ham} is the relativistic Dirac piece for electrons with Fermi velocity $v$ ($\approx 5\times 10^5$m/s for silicene\cite{Ezawa:2012a}).  Including the first two terms, one obtains the Kane-Mele Hamiltonian for intrinsic spin-orbit coupling\cite{Kane:2005} leading to a spin-orbit band gap of $\Delta_{\rm so}$.  The last term is associated with the on-site potential difference $\Delta_z$ which arises between the two sublattices when the system is subjected to an external electric field applied perpendicular to the plane of the lattice\cite{Ezawa:2012, Ezawa:2012a, Ezawa:2012b, Drummond:2012}.  As a matrix, Eqn.~\eqref{Ham} is,
\begin{align}\label{Ham-matrix}
\hat{H}_{\xi\sigma}=\left(\begin{array}{cc}
\Delta_{\xi\sigma} & \hbar v(\xi k_x-ik_y)\\
\hbar v(\xi k_x+ik_y) & -\Delta_{\xi\sigma}
\end{array}\right),
\end{align}
where $\Delta_{\xi\sigma}=(1/2)(-\xi\sigma\Delta_{\rm so}+\Delta_z)$, with $|\Delta_{++}|\equiv\Delta_{\rm min}$, and $|\Delta_{+-}|\equiv\Delta_{\rm max}$.  The eigenvalues of Eqn.~\eqref{Ham-matrix} are
\begin{align}
\varepsilon_{\xi\sigma}=\pm\sqrt{\hbar^2v^2k^2+\Delta_{\xi\sigma}^2}.
\end{align}
A schematic representation of the band structure at the $K$ point is given in Fig.~\ref{fig:Energy} for varying $\Delta_z$.  
Figure~\ref{fig:Energy}(a) shows the low-energy dispersion for the Kane-Mele QSHI ($\Delta_z=0$).  The two purple bands are spin degenerate, and separated by a gap of $\Delta_{\rm so}$.  For finite $\Delta_z<\Delta_{\rm so}$, the system remains a TI; however, the bands become spin split.  The gap of the spin-up band (blue) decreases while that of the spin-down band (red) increases [see Fig.~\ref{fig:Energy}(b)].  The critical value $\Delta_z=\Delta_{\rm so}$ is shown in Fig.~\ref{fig:Energy}(c).  Here, the lowest band gap closes forming a Dirac point while the upper gap has continued to increase.  For all $\Delta_z>\Delta_{\rm so}$, the system is a trivial BI [Fig.~\ref{fig:Energy}(d)].  With the reopening of the lowest gap, a band inversion occurs associated with a change in the pseudospin label of the two lowest gapped bands.  Now both energy gaps increase with $\Delta_z$, and the energy separation between the two spin bands at $k=0$ remains at $\Delta_{\rm so}$.  Throughout this paper, we will take $\Delta_{\rm so}=8$ meV.  At the $K^\prime$ point, identical behaviour is observed; however, the spin labels are interchanged.  Using a tight-binding hopping amplitude of $t=1.6$eV\cite{Liu:2011}, the low-energy approximation is quite accurate for energies between $\pm 800$ meV\cite{Ezawa:2012a}.

To discuss the magnetic properties of silicene, we must first consider the effect of an external magnetic field $B$ on the band structure.  By choice, we orient $B$ in the $z$ direction, and work in the Landau gauge.  Thus, the magnetic vector potential, given by $\bm{B}=\nabla\times\bm{A}$, is written as $\bm{A}=(-By,0,0)$.  Returning to Eqn.~\eqref{Ham}, the magnetic field changes the momentum operators by the usual Peierls substitution
\begin{align}
\hbar k_i\rightarrow\hbar k_i+eA_i,
\end{align}
where we work in units of $c=1$.  The low-energy Hamiltonian becomes
\begin{align}\label{Ham-B}
\hat{H}_{\xi\sigma}=\left(\begin{array}{cc}
\Delta_{\xi\sigma}-\frac{1}{2}g_s\mu_BB\sigma & \hbar v\left(\xi \left[k_x-\frac{eB}{\hbar}y\right]-ik_y\right)\\
\hbar v(\xi \left[k_x-\frac{eB}{\hbar}y\right]+ik_y) & -\Delta_{\xi\sigma}-\frac{1}{2}g_s\mu_BB\sigma
\end{array}\right),
\end{align}
where we include the Zeeman interaction $-(1/2)g_s\mu_BB\sigma_z$ in Eqn.~\eqref{Ham}; $g_s$ is the Zeeman coupling strength which we take to be\cite{Wang:2010} 23 to elucidate the interesting features; $\mu_B=e\hbar/(2m_e)\approx 5.78\times 10^{-2}$ meV/T is the Bohr magneton with $m_e$, the mass of an electron.  For silicene, the Zeeman energy is small, and usually ignored\cite{Tahir:2012a, Ezawa:2012, Ezawa:2012a, Ezawa:2012b, Tabert:2013a, Tabert:2013c}.  Retaining the Zeeman interaction, Eqn.~\eqref{Ham-B} can be solved to give the Landau level dispersion
\begin{align}\label{LL}
\mathcal{E}^{\xi\sigma}_{N,s}=\left\lbrace\begin{array}{cc}
-\frac{1}{2}g_s\mu_BB\sigma+s\sqrt{\Delta_{\xi\sigma}^2+2NE_1^2}, & N=1,2,3,...\\
-\xi\Delta_{\xi\sigma}-\frac{1}{2}g_s\mu_BB\sigma, & N=0
\end{array}\right.,
\end{align}
where $s=\pm$ is a band index, and $E_1\equiv\sqrt{\hbar ev^2B}\approx 12.82$ meV for $B=1$T.  Note: in a sum over $s=\pm$, there is only a single contribution from the $N=0$ levels.

\section{Magnetization and the Hall Effect}

\begin{figure}[h!]
\begin{center}
\includegraphics[width=1.0\linewidth]{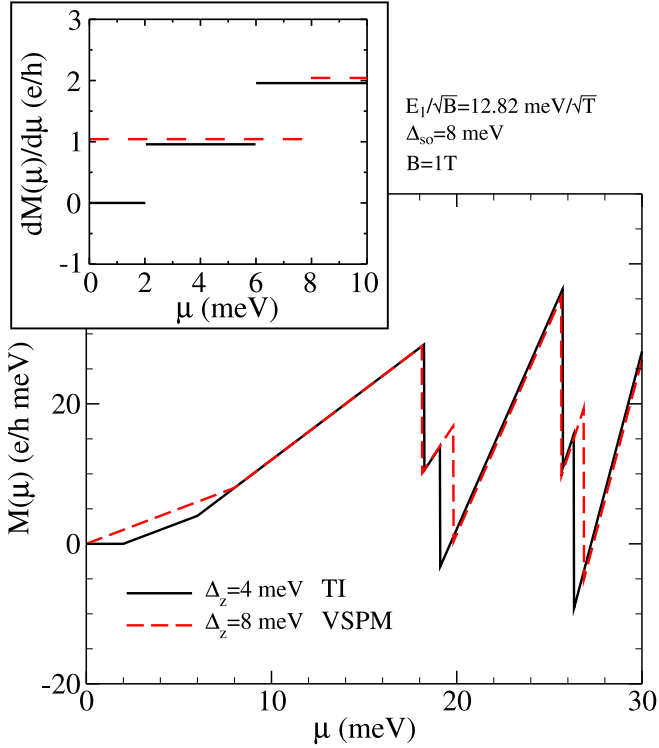}
\end{center}
\caption{\label{fig:Mag-Saw}(Color online) Magnetization as a function of $\mu$ in the TI phase (solid back curve), and VSPM regime (dashed red curve) for $g_s=0$.  Inset: the slope of the magnetization as a function of $\mu$ which is related to the Hall conductivity through $\partial M/\partial\mu=(1/e)\sigma_H$.  Note: in the TI regime, the Hall conductivity is zero for $\mu$ less than $\Delta_{\rm min}$.  In the VSPM, a finite $\sigma_H$ persists to $\mu\rightarrow 0$.  Plateaus have been offset from their integer values for clarity.
}
\end{figure}

\begin{figure}[h!]
\begin{center}
\includegraphics[width=1.0\linewidth]{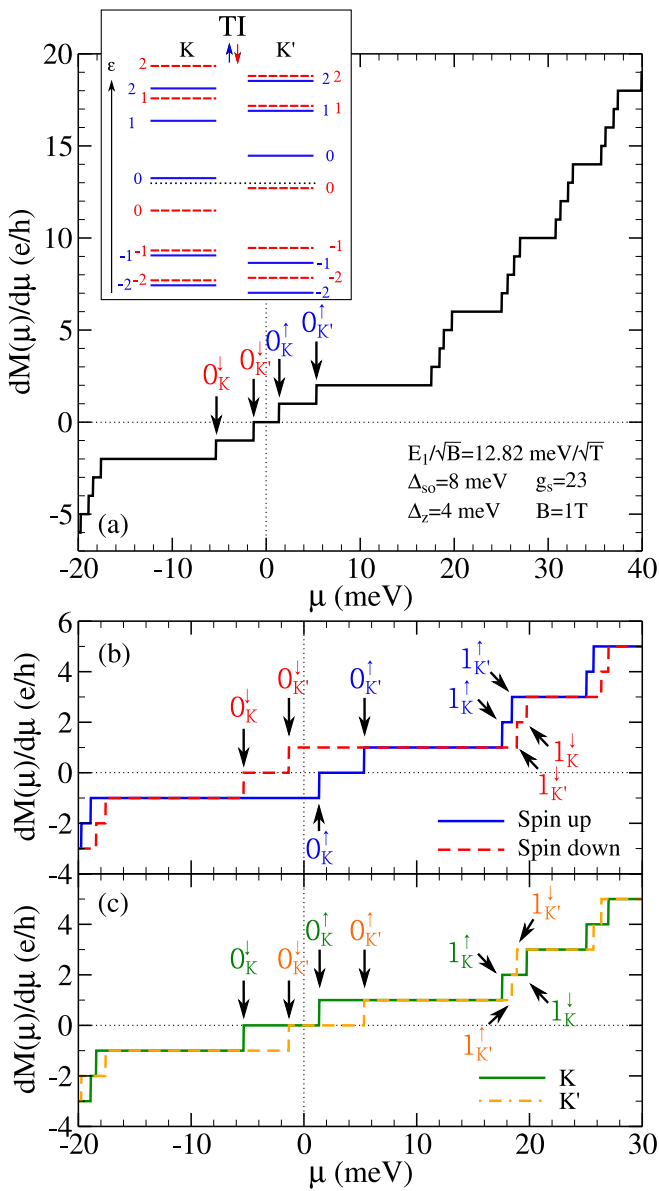}
\end{center}
\caption{\label{fig:Hall-TI}(Color online) The slope of the magnetization which gives the same integer values as the Hall conductivity.  (a) The charge Hall conductivity as a function of $\mu$ in the TI regime.  Four steps are seen near $\mu=0$ which are associated with the $N=0$ Landau levels.  A Zeeman interaction splits the two step feature of each of $N=1,2,3,..$ into a quartet of four plateaus.  The Landau levels at $K$, and $K^\prime$ are shown in the inset for finite $g_s$ .  Solid blue corresponds to spin-up, and dashed red to spin-down. (b) Spin contributions to the charge Hall conductivity. (c) Valley contributions to the Hall effect.
}
\end{figure}

\begin{figure}[h!]
\begin{center}
\includegraphics[width=0.86\linewidth]{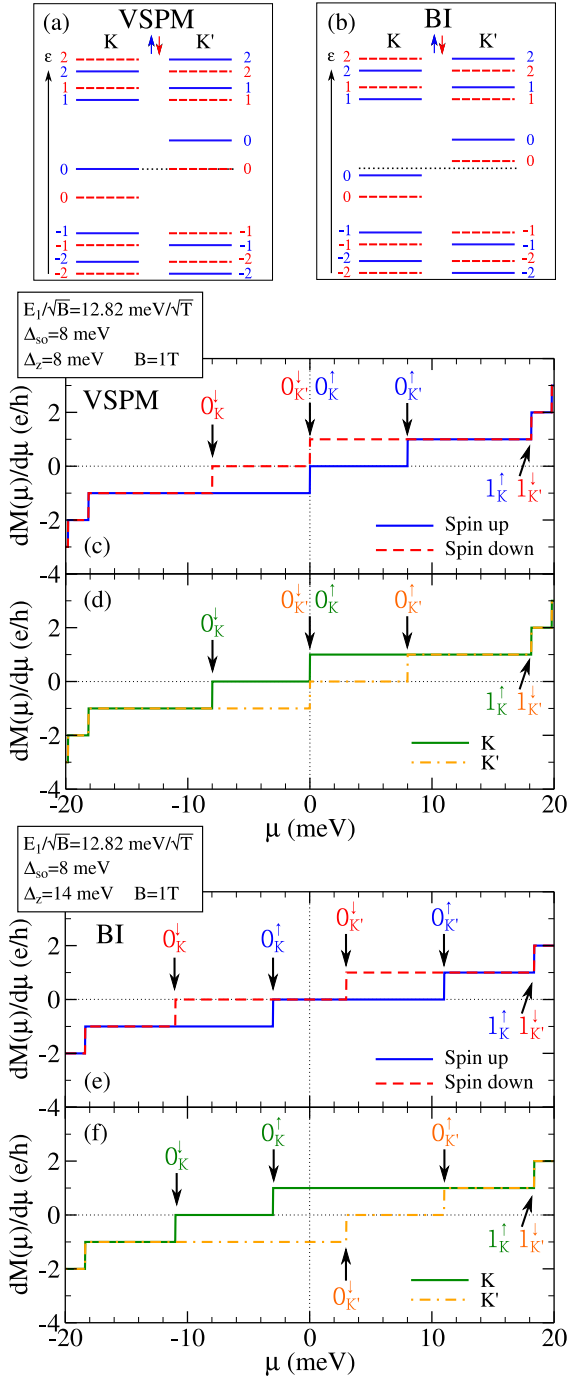}
\end{center}
\caption{\label{fig:Hall-BI-VSPM}(Color online) The valley separated Landau level spectrum for the (a) VSPM, and (b) BI regimes for $g_s=0$.  Solid blue lines represent spin-up, and dashed red correspond to spin-down. (c) Spin, and (d) valley contributions to the Hall conductivity for $\Delta_z=\Delta_{\rm so}$. Note that a valley-spin-polarized Hall effect persists to $\mu\rightarrow 0$.  (e) Spin, and (f) valley contributions to the Hall conductivity for the BI. Note that a valley-spin-polarized Hall effect is possible; however, there is no charge Hall response for $\mu\rightarrow 0$.
}
\end{figure}

Our examination of the magnetic response of silicene begins with the grand thermodynamic potential\cite{Sharapov:2004}
\begin{align}\label{Omega-T}
\Omega(T,\mu)=-T\int_{-\infty}^\infty N(\omega){\rm ln}\left(1+e^{(\mu-\omega)/T}\right)d\omega,
\end{align}  
where $T$ is the temperature, and $N(\omega)$ is the density of states.  Note that we have taken $k_B=1$.  In the absence of impurity scattering, the density of states in a magnetic field is given by a series of Dirac-delta functions located at the Landau level energies.  For a single $\xi$, and $\sigma$,
\begin{align}\label{DOS}
N_{\xi\sigma}(\omega)=\frac{eB}{h}\left[\delta\left(\omega-\mathcal{E}_0^{\xi\sigma}
\right)+\sum_{\substack{N=1 \\ s=\pm}}^\infty\delta\left(\omega-\mathcal{E}_{N,s}^{\xi\sigma}\right)\right].
\end{align}
As will be discussed further on, the quantities of interest depend on a derivative of the grand potential with respect to $B$; thus, for convenience, we choose to add $(\mu/2)\int_{-\infty}^\infty N(\omega)d\omega$ to Eqn.~\eqref{Omega-T}.  Since the integral of the density of states over all energies gives the total number of states (which is independent of $B$), this term will not contribute to the magnetization ($-\partial\Omega/\partial B$).  At zero temperature, Eqn.~\eqref{Omega-T} then becomes
\begin{align}
\Omega(\mu)=\int_{-\infty}^0\left(\omega-\frac{\mu }{2}\right)N(\omega)d\omega &+\int_0^\mu (\omega-\mu)N(\omega)d\omega\notag\\
&+\frac{\mu}{2}\int_0^\infty N(\omega)d\omega.
\end{align}
For silicene,
\begin{align}
\int_0^\infty N_{\xi\sigma}(\omega)d\omega=\int_{-\infty}^0 N_{\xi\sigma}(\omega)d\omega-\frac{eB}{h}\Upsilon,
\end{align}
where (for realistic values of Zeeman splitting)
\begin{align}
\Upsilon=\left\lbrace\begin{array}{cc}
-\sigma & \Delta_z<\Delta_{\rm so}\\
\displaystyle\frac{\xi-\sigma}{2} & \Delta_z=\Delta_{\rm so}, \, g_s=0\\
\xi & \Delta_z=\Delta_{\rm so}, \, g_s> 0\\
\xi & \Delta_z>\Delta_{\rm so}
\end{array}\right..
\end{align}
Therefore, 
\begin{align}\label{Omega-minus-vac}
\Omega_{\xi\sigma}(\mu)=\int_0^\mu (\omega-\mu)N_{\xi\sigma}(\omega)d\omega-\frac{eB\mu}{2h}\Upsilon+\int_{-\infty}^0\omega N_{\xi\sigma}(\omega)d\omega.
\end{align}
The final term (which does not depend on $\mu$) gives the vacuum contribution, and will simply provide a constant background to the $\mu$ dependence of the magnetization.    In the absence of impurity scattering, when summed over $\xi$ and $\sigma$, the first two terms of Eqn.~\eqref{Omega-minus-vac} give
\begin{align}\label{Omega-tilde}
\tilde{\Omega}(\mu)=\frac{eB}{h}&\sum_{\xi,\sigma=\pm} \bigg[\left(\mathcal{E}_0^{\xi\sigma}-\mu\right)\Theta\left(\mu-\mathcal{E}_0^{\xi\sigma}\right)\Theta\left(\mathcal{E}_0^{\xi\sigma}\right)\notag\\
&\left.-\frac{\mu}{2}\Upsilon+\sum_{N=1}^\infty\left(\mathcal{E}^{\xi\sigma}_{N,+}-\mu\right)\Theta\left(\mu-\mathcal{E}^{\xi\sigma}_{N,+}\right)\right],
\end{align}
where we have used Eqn.~\eqref{DOS} for the density of states, and have assumed that all the $s=-$ states are negative.  It is important to note that $\Theta(0)\equiv 1/2$ as only half the delta function situated at $\omega=0$ is integrated.  A plot of the magnetization derived from Eqn.~\eqref{Omega-tilde} is shown in Fig.~\ref{fig:Mag-Saw} for $\Delta_z=4$ (solid black), and 8 meV (dashed red).  
In both cases, a jagged saw-tooth oscillation is seen.  For $\Delta_z=4$ meV, the system is in the TI phase and, thus, two $N=0$ Landau levels are at positive energy.  For low chemical potential, two kinks are seen in $M(\mu)$ at $\mu=2$, and 6 meV associated with the spin-up $N=0$ levels at $K$, and $K^\prime$, respectively.  For the VSPM, only one Landau level exists for $\omega>0$ and, hence, only one kink is observed at $\mu=8$ meV.  When $\mu$ is greater than the $N=0$ levels, the usual saw-tooth behaviour is present.  The jumps occur at the Landau level energies of all the $N>0$ levels.  Two teeth exist close in energy due to the spin-splitting of the levels (see the insets of Figs.~\ref{fig:Hall-TI}, and \ref{fig:Hall-BI-VSPM}).  The magnetization in the BI regime is similar to that of the TI; however, the two low-energy kinks are associated with the spin-down, and -up $N=0$ levels at $K^\prime$.

The slope of the magnetization is of interest as it is related to the robust quantization of the Hall conductivity through the Streda formula\cite{Smrcka:1977} $\partial M(\mu)/\partial\mu=(1/e)\sigma_H$.  The slope is shown in the inset of Fig.~\ref{fig:Mag-Saw}.  For the TI, no Landau levels exist at zero energy, and thus the Hall conductivity is zero until $\mu$ reaches the energy of the lowest level.  At this energy, it steps up by one unit of $e^2/h$ until the next level at which point it increments by another $e^2/h$.  For the VSPM, there does exist a level at charge neutrality, and thus the slope of the magnetization is finite for zero chemical potential.  The BI behaves similar to the TI.

The quantization of $\partial M/\partial\mu$ can be seen by taking the $\mu$ derivative of Eqn.~\eqref{Omega-tilde}.  For $\mu>0$, we find
\begin{align}\label{Omega-dmu}
\frac{\partial\tilde{\Omega}}{\partial\mu}=\frac{eB}{h}&\sum_{\xi,\sigma=\pm} \left[-\Theta\left(\mu-\mathcal{E}_0^{\xi\sigma}\right)\Theta\left(\mathcal{E}_0^{\xi\sigma}\right)-\frac{\Upsilon}{2}\right.\notag\\
&\left.-\sum_{N=1}^\infty\Theta\left(\mu-\mathcal{E}^{\xi\sigma}_{N,+}\right)\right],
\end{align}
and, the slope of the magnetization is then
\begin{align}\label{Mag-dmu}
-\frac{\partial}{\partial B}\frac{\partial\tilde{\Omega}}{\partial\mu}\equiv \frac{\partial M}{\partial\mu}=&\frac{e}{h} \sum_{\xi,\sigma=\pm} \bigg[\Theta\left(\mu-\mathcal{E}_0^{\xi\sigma}\right)\Theta\left(\mathcal{E}_0^{\xi\sigma}\right)+\frac{\Upsilon}{2}\notag\\
&\left. +\sum_{N=1}^\infty\Theta\left(\mu-\mathcal{E}^{\xi\sigma}_{N,+}\right)\right].
\end{align}
For the TI, and BI regimes in the absence of Zeeman splitting, the Hall conductivity $\sigma_H=(e^2/h)\nu$ has filling factors $\nu=0,\pm 1, \pm 2, \pm 4, \pm 6,...$; while, the VSPM has filling factors $\nu=\pm 1,\pm 2, \pm 4, \pm 6,...$.  If $\Delta_z=0$, the familiar values of gapped graphene ($\nu=0, \pm 2, \pm 6, \pm 10...$) are retained.  For $\Delta_{\rm so}=0$, and $\Delta_z=0$, the famous graphene values $\nu=\pm 2, \pm 6, \pm 10,...$ are observed.

Zeeman splitting breaks the valley degeneracy of the $N>0$ Landau levels, and the Hall conductivity can take any integer value of $e^2/h$.
Figure~\ref{fig:Hall-TI}(a) shows the slope of $M(\mu)$ in the presence of Zeeman interactions.  The steps associated with the $N=0$ levels are shifted [higher(lower) in $\mu$ for spin-up(down)].  The steps at higher $\mu$ are split into four integer plateaus which are close in energy.  For $g_s=0$, the four steps would reduce to two steps of height $2e^2/h$.  The inset shows a schematic of the Landau level energies for a TI at the two valleys $K$, and $K^\prime$.  The solid blue lines represent spin-up while the dashed red lines correspond to the spin-down levels.  $\mu=0$ is given by the dotted black line.  The effect of Zeeman splitting in gapped graphene has previously been examined\cite{Gusynin:2006b}.  For graphene with gap $\Delta$, there are two spin-degenerate $N=0$ Landau levels at $-\xi\Delta$.  Zeeman interactions split the degeneracy, and four $N=0$ plateaus are seen\cite{Gusynin:2006b}.  The four-fold valley-spin-degenerate $N>0$ levels become spin-split, and a double step structure is observed\cite{Gusynin:2006b}.  This is equivalent to the limit $\Delta_{\rm so}=0$, and $\Delta_z/2=\Delta$.  Note that these results are for $T=0$.  A finite temperature will smear the edges of the steps on an energy scale associated with $k_BT$, but will leave the quantization unaffected.  Therefore, a very low temperature is required to see the fine structure, due to the Zeeman interaction, which is seen above the central quartet of plateaus.

Figure~\ref{fig:Hall-TI}(b) shows the spin contributions to the Hall conductivity [solid blue(dashed red) for spin-up(down)].  A negative Hall conductivity corresponds to a net edge current in the opposite direction to that of the positive Hall effect.  Thus, while at $\mu=0$ the charge Hall conductivity is zero, there is a net spin Hall effect with a spin-up conductance of $2e^2/h$ in the negative direction.  For $\mu$ between the two spin-up $N=0$ Landau levels, only the spin-down electrons contribute to the Hall effect.  When $\mu$ is between the two spin-down $N=0$ levels, only the spin-up electrons contribute.  We note that for finite Zeeman splitting, the quartet of steps associated with each of the $N=1,2,3,...$ levels also provides a spin imbalance.  While the $\nu=\pm 2$ Hall effect is spin degenerate, the first two steps of the $N=1$ levels provide a spin-polarized response.  This is lost when $\mu$ becomes greater than the remaining $N=1$ levels.  The pattern continues for higher values of $N$.  Without a Zeeman interaction, all the $N\neq 0$ steps are spin degenerate.   

Figure~\ref{fig:Hall-TI}(c) shows the valley contributions to the Hall effect (solid green for $K$, and dashed orange for $K^\prime$).  At charge neutrality, there is no net valley conductivity.  For $\mu$ between the two positive $N=0$ Landau levels, only the $K$ point contributes; while, for $\mu$ between the two negative $N=0$ levels, only $K^\prime$ contributes.  Combining this result with those of Fig.~\ref{fig:Hall-TI}(b), we note that for $\mu$ between the two spin-up $N=0$ Landau levels, the finite charge Hall effect is accompanied by finite spin-down, and momentum $K$ spin, and valley Hall effects, respectively.  Likewise, for $\mu$ between the two spin-down $N=0$ levels, there is a finite negative spin-up, and momentum $K^\prime$ spin, and valley Hall effect in addition to the charge Hall conductivity.   Again, we note that Zeeman splitting allows for a valley polarized response from the $N>0$ levels.

The spin, and valley contributions to the Hall effect in the VSPM, and BI regimes for $g_s=0$ are given by Figs.~\ref{fig:Hall-BI-VSPM}(c), (d) and (e), (f), respectively.  Schematic representations of the Landau levels in these regimes can be seen in Figs.~\ref{fig:Hall-BI-VSPM}(a), and (b), respectively.  Again, solid blue lines represent spin-up, and dashed red lines correspond to spin-down.
The charge Hall effect, is given by the sum of the individual spin/valley conductivities.  Therefore, in the VSPM phase, a finite charge Hall effect is present at $\mu=0$; like the TI, there is no charge Hall conductivity at $\mu=0$ for the BI.  The spin, and valley contributions to the VSPM Hall effect [see Figs.~\ref{fig:Hall-BI-VSPM}(c), and (d), respectively] reveal that for $0<\mu<\mathcal{E}_{0}^{K^\prime\uparrow}$, the finite charge Hall conductivity is comprised of spin-down, and momentum $K$ fermions.  For $\mathcal{E}_{0}^{K\downarrow}<\mu<0$, the charge Hall conductivity is made of spin-up, and momentum $K^\prime$ fermions.  In the BI regime [see Figs.~\ref{fig:Hall-BI-VSPM}(e), and (f)], while there is no charge Hall effect for $\mu=0$, there is a finite valley Hall effect for $\mathcal{E}_{0}^{K\uparrow}<\mu<\mathcal{E}_{0}^{K^\prime\downarrow}$.  For $\mathcal{E}_{0}^{K^\prime\downarrow}<\mu<\mathcal{E}_{0}^{K^\prime\uparrow}$, the charge Hall conductivity is made of spin-down, and momentum $K$ electrons.  For negative $\mu$ between $\mathcal{E}_{0}^{K\downarrow}$, and $\mathcal{E}_{0}^{K\uparrow}$, the Hall effect is comprised of spin-up, and momentum $K^\prime$ electrons. 

To summarize, the total Hall conductivity at small $\mu$ is zero for the TI, and BI; however, a finite spin Hall effect is present in the TI regime while an analogous valley Hall effect is found in the BI regime.  The finite charge Hall conductivity is also accompanied by spin-valley Hall effects for various values of $\mu$.  Including Zeeman interactions results in a four step feature associated with each of the $N>0$ levels.  This also permits a valley-spin imbalance.  By contrast, when $|\mu|$ is small in the VSPM phase, the total Hall conductivity is finite, and carries the sign of the chemical potential.  Spin, and valley Hall effects are also seen.  Realistically, the Zeeman interaction will shift the Landau levels by much less than the inter-level spacing\cite{Tabert:2013c}.  Therefore, in what follows, we ignore the Zeeman term as its effects are negligible.

As the filling factors, and valley contributions to the Hall effect calculated from the magnetization are in contrast to what is predicted by Tahir and Schwingenschl{\"o}gl for the DC Hall conductivity\cite{Tahir:2013}, we have verified our results by direct calculation of $\sigma_{xy}(\Omega\rightarrow 0)$ given by Eqn.~(14) in Ref.~\cite{Tabert:2013c}.  Reference~\cite{Tabert:2013c} used the Kubo formula and, thus, represents an independent check.  In the DC limit $\Omega\rightarrow 0$ for $\mu>0$, we find\cite{Tabert:2013c}

\begin{align}\label{Condxy-Re}
\sigma_{xy}(\mu)&=-\frac{e^2}{h}E_1^2\sum_{\xi, \sigma=\pm}\sum_{\substack{n,m=0 \\ s=\pm}}^\infty\xi\frac{\Theta\left(\mu-\mathcal{E}_{m,s}^{\xi\sigma}\right)-\Theta\left(\mu-\mathcal{E}_{n,+}^{\xi\sigma}\right)}{\left(\mathcal{E}_{n,+}^{\xi\sigma}-\mathcal{E}_{m,s}^{\xi\sigma}\right)^2}\notag\\
&\times\left[\left(A_{m,s}B_{n,+}\right)^2\delta_{n,m-\xi}
-\left(B_{m,s}A_{n,+}\right)^2\delta_{n,m+\xi}\right],
\end{align}
where\cite{Tabert:2013a, Tabert:2013c}
\begin{equation}\label{An}
A_{n,s}=\left\lbrace\begin{array}{cc}
\displaystyle\frac{s\sqrt{|\mathcal{E}_{n,s}^{\xi\sigma}|+s\Delta_{\xi\sigma}}}{\sqrt{2|\mathcal{E}_{n,s}^{\xi\sigma}|}}, &\quad n\neq 0,\\
\displaystyle\frac{1-\xi}{2}, &\quad n=0,
\end{array}\right.
\end{equation}
and
\begin{equation}\label{Bn}
B_{n,s}=\left\lbrace\begin{array}{cc}
\displaystyle\frac{\sqrt{|\mathcal{E}_{n,s}^{\xi\sigma}|-s\Delta_{\xi\sigma}}}{\sqrt{2|\mathcal{E}_{n,s}^{\xi\sigma}|}}, &\quad n\neq 0,\\
\displaystyle\frac{1+\xi}{2}, &\quad n=0,
\end{array}\right.
\end{equation}
which gives the same quantized plateaus as Eqn.~\eqref{Mag-dmu} after multiplication by a factor of $e$.

\subsection{Integrated Density of States and Optical Spectral Weight}

\begin{figure}[h!]
\begin{center}
\includegraphics[width=1.0\linewidth]{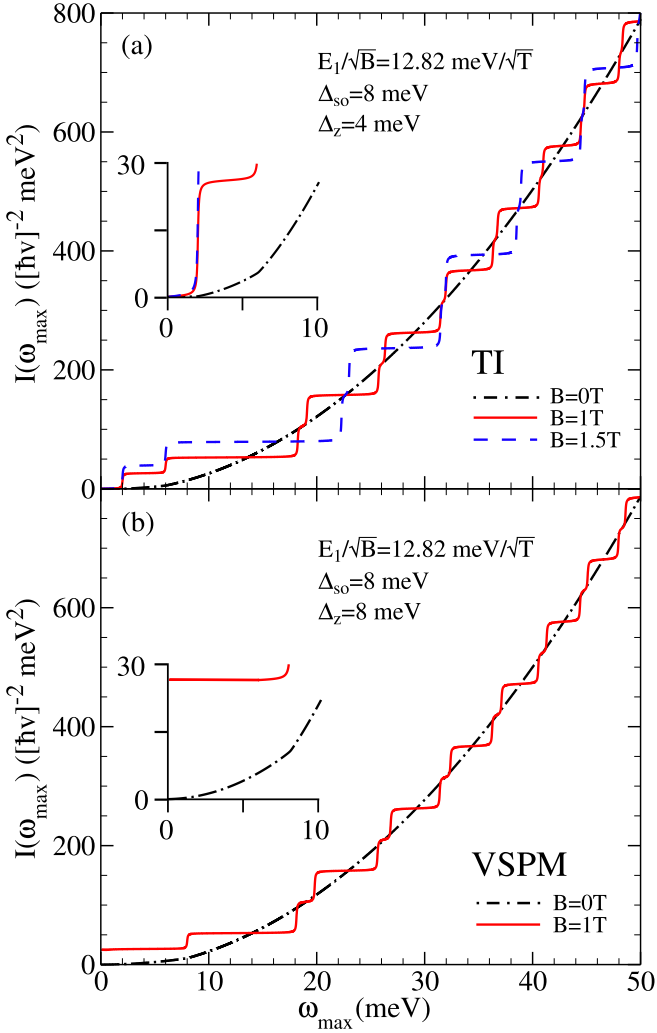}
\end{center}
\caption{\label{fig:DOS-Int}(Color online) (a) Integrated density of states for finite $\Delta_z<\Delta_{\rm so}$ and $B$=0 (dash-dotted black curve), $B=1$T (solid red), and $B=1.5$T (dashed blue). For $B=0$, the integrated density of states has kinks at the values of the gaps (emphasized in the inset). For $\omega_{\rm max}<\Delta_{\rm min}$, $I(\omega_{\rm max})=0$.  (b) For $\Delta_{\rm so}=\Delta_z$, a finite $I(\omega_{\rm max})$ exists for all $\omega_{\rm max}$ due to Landau levels at zero energy.
}
\end{figure}

A similar quantization of plateaus is obtained for the integrated density of states.  We define the total integrated density of states up to energy $\omega_{\rm max}$ as
\begin{align}\label{DOS-Int}
I(\omega_{\rm max})=\sum_{\xi,\sigma=\pm}\int_0^{\omega_{\rm max}}N_{\xi\sigma}(\omega)d\omega,
\end{align}
where $N_{\xi\sigma}(\omega)$ is given by Eqn.~\eqref{DOS}. In Fig.~\ref{fig:DOS-Int}, $I(\omega_{\rm max})$ is plotted as a function of the cutoff energy $\omega_{\rm max}$.  
Figure~\ref{fig:DOS-Int}(a) applies to the TI phase for which the two gaps $\Delta_{\rm min}$ and $\Delta_{\rm max}$ are both finite, and equal to 2 and 6 meV, respectively.  Three values of magnetic field are presented: $B=1$T (solid red curve), $B=1.5$T (dashed blue curve), and $B=0$T (dash-dotted black curve).  We begin our discussion with $B=0$T.  Since all the electronic states are gapped, $I(\omega_{\rm max})=0$ for $\omega_{\rm max}<\Delta_{\rm min}$=2 meV.  There is a change in slope at $\omega_{\rm max}=\Delta_{\rm max}=6$ meV which is highlighted in the inset.  At higher energy, $I(\omega_{\rm max})$ is proportional to $\omega_{\rm max}^2$ which is traced back to the linear energy dispersion of the Dirac fermions.  When the magnetic field is finite, the density of states forms a series of Landau level peaks but $I(\omega_{\rm max})$ remains zero for $\omega_{\rm max}<\Delta_{\rm min}=2$ meV after which it has a vertical step associated with the occupation of the lowest Landau level.  The height of the curve reflects the spectral weight of the delta function.  The first two steps have the same height, and the other steps are twice as large due to the spin, and valley degeneracy of the $N\neq 0$ levels.  Note that the energy range between the steps is alternately small, and large, and that $I(\omega_{\rm max})$ tends toward its $B=0$ value as $\omega_{\rm max}$ increases.  For $B=1.5$T, the height of the steps has increased due to the $B$ dependence of the Landau level degeneracy [see Eqn.~\eqref{DOS}].  The energy range over which a given step persists is also changed as it reflects the energy between adjacent Landau levels.  Similar results are found for the BI regime.  In Fig.~\ref{fig:DOS-Int}(b), we show equivalent results for the VSPM case which corresponds to $\Delta_z=\Delta_{\rm so}$.  Here, $\Delta_{\rm min}=0$, and $\Delta_{\rm max}=8$ meV.  When $B=0$T (dash-dotted black curve), a change in slope is seen for $I(\omega_{\rm max})$ at $\omega_{\rm max}=\Delta_{\rm max}$.  However, a finite integrated density of states persists to the lowest energy $\omega_{\rm max}\rightarrow 0$.  In fact, in a finite magnetic field, $I(\omega_{\rm max}\rightarrow 0)\neq 0$, and is constant in the entire range of $\omega_{\rm max}<\Delta_{\rm max}$.   This strikingly different behaviour between the VSPM, and TI/BI regimes, seen in both the Hall conductivity, and integrated density of states, can be employed to monitor the phase transition between the two states of matter.

\begin{figure}[h!]
\begin{center}
\includegraphics[width=.9\linewidth]{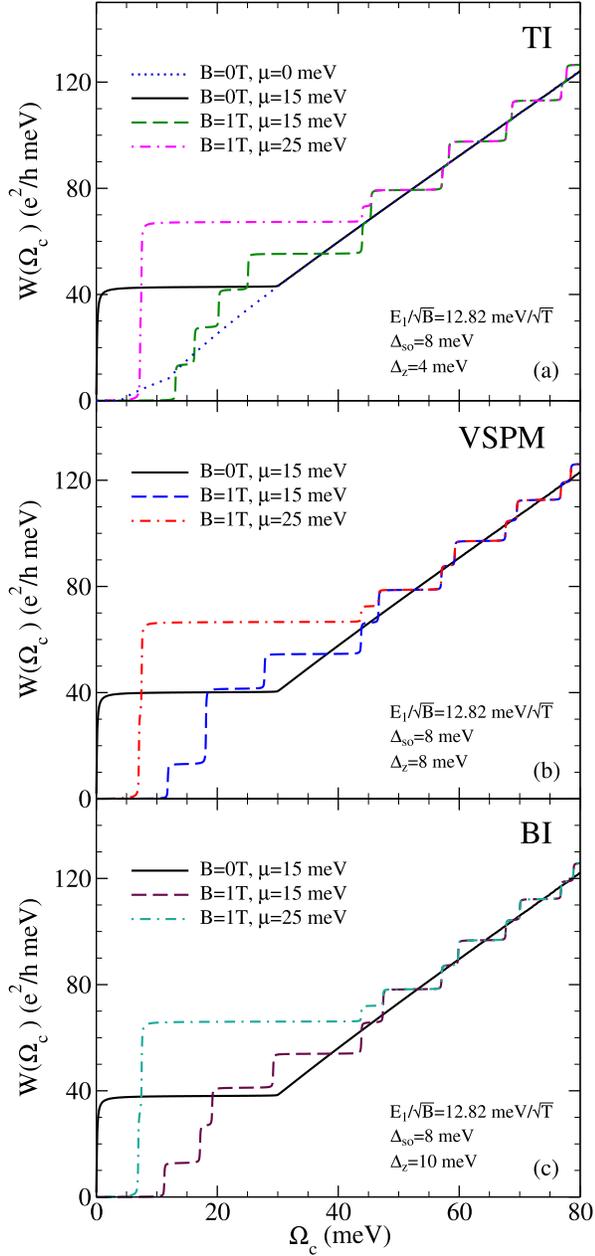}
\end{center}
\caption{\label{fig:OSW}(Color online) Optical spectral weight in the (a) TI, (b) VSPM, and (c) BI phases.  (a) For $\mu$ between the $N=0$, and 1 levels (dashed green curve), there are four low-energy steps. (b) For the VSPM, only three steps result from the $N=0$ to 1 optical transitions (dashed blue curve).  (c) The four low-energy steps reappear in the BI regime (dashed purple curve). 
}
\end{figure}

Plateaus similar to those described in the integrated density of states also exist for the absorptive part of the AC longitudinal conductivity integrated to $\Omega_c$.  The real part of $\sigma_{xx}(\Omega)$ is given by Eqn.~(12) of Ref.~\cite{Tabert:2013c}; the optical spectral weight up to $\Omega_c$ (denoted by $W(\Omega_c)=\int_{0}^{\Omega_c}\sigma_{xx}(\Omega)d\Omega)$ is
\begin{align}\label{OSW}
W(\Omega_c)&=\frac{e^2}{h}\pi E_1^2\sum_{\xi, \sigma=\pm}\sum_{\substack{n,m=0 \\ s=\pm}}^\infty\frac{\Theta\left(\mu-\mathcal{E}_{m,s}^{\xi\sigma}\right)-\Theta\left(\mu-\mathcal{E}_{n,+}^{\xi\sigma}\right)}{\mathcal{E}_{n,+}^{\xi\sigma}-\mathcal{E}_{m,s}^{\xi\sigma}}\notag\\
&\times\left[\left(A_{m,s}B_{n,+}\right)^2\delta_{n,m-\xi}
+\left(B_{m,s}A_{n,+}\right)^2\delta_{n,m+\xi}\right]\notag\\
&\times\Theta\left(\hbar\Omega_c+\mathcal{E}_{m,s}^{\xi\sigma}-\mathcal{E}_{n,+}^{\xi\sigma}\right)\Theta\left(\mathcal{E}_{n,+}^{\xi\sigma}-\mathcal{E}_{m,s}^{\xi\sigma}\right).
\end{align}
Equation~\eqref{OSW} [like Eqn.~\eqref{Condxy-Re}], is a simple algebraic sum that depends only on the Landau level energies $\mathcal{E}_{n,s}^{\xi\sigma}$, and gaps $\Delta_{\xi\sigma}$.  It is readily evaluated.   Results for the steps in the integrated optical conductivity are presented in the three frames of Fig.~\ref{fig:OSW} for the TI, VSPM, and BI.  
These results were obtained in two independent ways: by direct integration of the AC conductivity, and by employing Eqn.~\eqref{OSW}.  Figure~\ref{fig:OSW}(a), in which $\Delta_z=4$ meV, shows results for the TI side of the VSPM transition point.  The dashed green curve is for a magnetic field of 1T with chemical potential $\mu=15$ meV.  The dash-dotted magenta curve corresponds to $\mu=25$ meV, and $B=1$T.  For $\mu=15$ meV, and finite $B$, the chemical potential is above all the $N=0$ Landau levels but below all the $N=1$ levels.  The first four plateaus come from the four lowest optical transitions that are possible; these involve only $N=0$ to 1 transitions (see Refs.~\cite{Tabert:2013a,Tabert:2013c}).  For larger values of $\Omega_c$, additional optical excitations are possible.  These involve higher Landau levels within the constraint of the optical selection rules ($|N|\rightarrow|N|\pm 1$) which are built into Eqn.~\eqref{OSW} through the Kronecker deltas.  While Eqn.~\eqref{OSW} is compact, it is worth while writing down a more explicit form for the particular case of the dashed green curve of Fig.~\ref{fig:OSW}.  It is
\begin{align}\label{OSW-app}
W(\Omega_c)&=\frac{e^2}{h}\frac{\pi E_1}{2}\sum_{\xi, \sigma=\pm}\sum_{n=0}^{\infty}\Theta\left(\Omega_c-\mathcal{E}^{\xi\sigma}_{n+1,+}+\mathcal{E}^{\xi\sigma}_{n,-}\right)\notag\\
&\times\left(1+\frac{\bar{\Delta}_{\xi\sigma}^2}{\sqrt{2(n+1)+\bar{\Delta}_{\xi\sigma}^2}\sqrt{2n+\bar{\Delta}_{\xi\sigma}^2}}\right)\notag\\
&\times\frac{\Theta\left(\mathcal{E}^{\xi\sigma}_{n+1,+}-\mu\right)\Theta\left(\mu-\mathcal{E}^{\xi\sigma}_{n,-}\right)}{\sqrt{2n+\bar{\Delta}_{\xi\sigma}^2}+\sqrt{2(n+1)+\bar{\Delta}_{\xi\sigma}^2}}.
\end{align}  
Again, all plateaus in $W(\Omega_c)$ depend only on the energies $\mathcal{E}_{n,s}^{\xi\sigma}$, and the gaps $\Delta_{\xi\sigma}$ which appear normalized ($\bar{\Delta}_{\xi\sigma}\equiv \Delta_{\xi\sigma}/E_1$) in Eqn.~\eqref{OSW-app}.  The solid black curve of Fig.~\ref{fig:OSW}(a) is for no magnetic field\cite{Stille:2012}.  This curve shows no quantized plateaus but provides an unstructured background on which the discreteness of the Landau levels superimpose a step structure.  Importantly, as $\Omega_c$ is increased, the signature of the Landau level quantization becomes smaller, and fades into the background.  This is irrespective of $B$ or $\mu$ as illustrated by the dash-dotted magenta curve.  Returning to the $B=0$ case, the first step in the solid black curve gives the optical spectral weight of the Drude contribution to the conductivity\cite{Gusynin:2006a,Gusynin:2007,Gusynin:2007b,Gusynin:2009,Stille:2012}
\begin{align}
W_{\rm D}=\sum_{\xi,\sigma=\pm}\frac{e^2\pi}{4h}\frac{\mu^2-\Delta_{\xi\sigma}^2}{|\mu|}\Theta\left(|\mu|-|\Delta_{\xi\sigma}|\right).
\end{align} 
For $\Omega_c>2\mu>2|\Delta_{\xi\sigma}|$, the nearly linear slope above the Drude plateau is due to the universal background\cite{Ando:2002, Gusynin:2006a, Gusynin:2007, Gusynin:2007b, Gusynin:2009, Carbotte:2010, Stille:2012} in the optical conductivity.  This provides a spectral weight of
\begin{align}
W_{\rm BG}=\sum_{\xi,\sigma=\pm}\frac{e^2\pi}{8h}\left[\Omega_c-2\mu+4\Delta_{\xi\sigma}^2\left(\frac{1}{2\mu}-\frac{1}{\Omega_c}\right)\right]
\end{align}
which is a linear function of $\Omega_c$ when $\Omega_c\gg\mu$ and $|\Delta_{\xi\sigma}|$.  The dotted blue curve of Fig.~\ref{fig:OSW}(a) is also for $B=0$ but here the chemical potential is set at $\mu=0$.  Therefore, $W_{\rm D}=0,$
and
\begin{align}
W_{\rm BG}=\sum_{\xi,\sigma=\pm}\frac{e^2\pi}{8h}\left[\Omega_c-\frac{4\Delta_{\xi\sigma}^2}{\Omega_c}\right]\Theta\left(\Omega_c-2|\Delta_{\xi\sigma}|\right),
\end{align}
which has the same slope as the solid black curve for $\Omega_c\rightarrow\infty$, and merges with it above 30 meV.  However, for small $\Omega_c$, the dotted blue curve goes to zero since $\mu$ is below both gaps, and there is no Drude response.

Figure~\ref{fig:OSW}(b) is for the VSPM.  This is the special case which exists at the boundary between the TI [Fig.~\ref{fig:OSW}(a)], and the BI [Fig.~\ref{fig:OSW}(c)].  In this phase, the minimum gap is zero, and there are only three steps (as opposed to four) associated with the lowest energy optical transitions which involve the $N=0$ levels (dashed blue curve).  This is a distinct feature of this particular case.  As seen in Fig.~\ref{fig:OSW}(c), the four steps are restored in the BI regime.  When comparing the optical spectral weight with the integrated density of states in Fig.~\ref{fig:DOS-Int}, it is important to realize that the energies at which a new step appears is set by the difference between two Landau levels for which an optical transition is allowed.  For the density of states, and the Hall plateaus, the energies of the steps are set by the Landau level energies themselves.  A detailed discussion of the magneto-optical, and optical conductivity of silicene is found in Refs.~\cite{Tabert:2013a,Tabert:2013c}, and Ref.~\cite{Stille:2012}, respectively.  Closely related work\cite{ZLi:2013,ZLi:2014} includes a quadratic-in-momentum (Schr{\"o}dinger) contribution in addition to the linear-in-momentum Dirac term of Eqn.~\eqref{Ham} in the absence of buckling.  Including impurity effects will cause the edge of the steps to broaden and smear over an energy range of order $\Gamma$\cite{ZLi:2014}.  The predicted step structure will therefore be resolved for $\Gamma$ much less than the Landau level spacing (for the Hall conductivity and integrated density of states), or the allowed optical transition energies (for the optical spectral weight).  While these energy scales depend on $B$, for the moderate magnetic fields used herein, the dominant steps should be visible for $\Gamma\sim\mathcal{O}({\rm 1 meV})$. 

\section{Magnetization and Magnetic Susceptibility}

\subsection{Zero Temperature, Clean Limit}
\begin{figure}[h!]
\begin{center}
\includegraphics[width=1.0\linewidth]{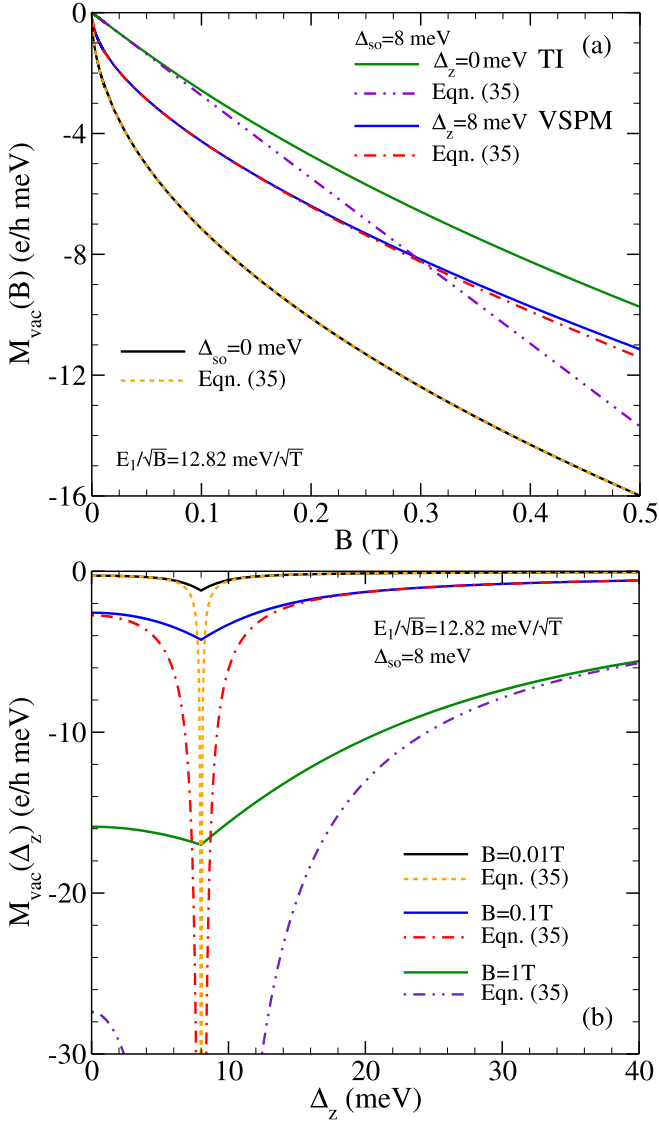}
\end{center}
\caption{\label{fig:Mag-Vac}(Color online) Vacuum contribution to the magnetization as a function of (a) $B$, and (b) $\Delta_z$.  (a) The agreement between Eqn.~\eqref{Mag-vac} (solid curves), and Eqn.~\eqref{Mag-approx} is good for small $B$ or $\Delta_{\rm so}=\Delta_z=0$.  (b) Here, agreement is found between the full result (solid curves), and approximate expression when $|\Delta_z|\gg\Delta_{\rm so}$ or $\ll\Delta_{\rm so}$, and $B$ is small.  A minimum is seen in $M_{\rm vac}$ in the VSPM phase.  
}
\end{figure}

For convenience, we now choose to work with the relativistic form of the grand potential.  That is\cite{Sharapov:2004},
\begin{align}\label{Omega-Rel-T}
\Omega(T,\mu)=-T\int_{-\infty}^\infty N(\omega){\rm ln}\left(2{\rm cosh}\frac{\omega-\mu }{2T}\right)d\omega.
\end{align}
For our purposes, this form is equivalent to Eqn.~\eqref{Omega-T}\cite{Tabert:2015}.  Equation~\eqref{Omega-Rel-T} has been worked out analytically for gapped graphene at $T=0$\cite{Sharapov:2004}, and may be extended to silicene.  After applying the Poisson summation formula, the grand potential may be written as the sum of a regular, vacuum, and oscillating piece.  For $\mu\geq 0$, the single spin, and valley contributions are
\begin{align}\label{Omega-reg}
\Omega^{\xi\sigma}_{\rm reg}(\mu)&=-\frac{eB}{2h}\Theta\left(\mu-|\Delta_{\xi\sigma}|\right)\left[\frac{\mu(\mu^2-3\Delta_{\xi\sigma}^2)}{3E_1^2}-|\Delta_{\xi\sigma}|\right.\notag\\
&\left.-2^{3/2}E_1\zeta\left(-\frac{1}{2}, 1+\frac{\Delta_{\xi\sigma}^2}{2E_1^2}\right)\right],
\end{align}
\begin{align}\label{Omega-vac}
\Omega^{\xi\sigma}_{\rm vac}(\mu=0)&=-\frac{eB}{2h}\left[\frac{\Lambda\Delta_{\xi\sigma}^2}{\sqrt{\pi}E_1^2}+|\Delta_{\xi\sigma}|\right.\notag\\
&\left.+2^{3/2}E_1\zeta\left(-\frac{1}{2}, 1+\frac{\Delta_{\xi\sigma}^2}{2E_1^2}\right)\right]+\mathcal{O}\left(\frac{1}{\Lambda}\right),
\end{align}
and
\begin{align}\label{Omega-osc}
\Omega^{\xi\sigma}_{\rm osc}(\mu)=\frac{eB}{2h}\frac{E_1^2}{\mu}\sum_{k=1}^\infty\frac{1}{(\pi k)^2}{\rm cos}\left(\frac{\pi k(\mu^2-\Delta_{\xi\sigma}^2)}{E_1^2}\right),
\end{align}
respectively [see Eqns. (6.3), (A4), and (8.7) in Ref.\cite{Sharapov:2004}], where $\zeta(-1/2,x)$ is the Hurwitz zeta function, and $\Lambda$ is an ultraviolet cutoff associated with the band width.  The magnetization is the sum of
\begin{align}\label{Mag-reg}
M_{\rm reg}(\mu)&=-\frac{e}{h}\sum_{\xi,\sigma=\pm}\Theta\left(\mu-|\Delta_{\xi\sigma}|\right)\mathcal{C}(\Delta_{\xi\sigma},B),
\end{align}
\begin{align}\label{Mag-vac}
M_{\rm vac}(\mu=0)&=\frac{e}{h}\sum_{\xi,\sigma=\pm}\mathcal{C}(\Delta_{\xi\sigma},B),
\end{align}
and
\begin{align}\label{Mag-osc}
M_{\rm osc}(\mu)=-\frac{e}{h}\frac{E_1^2}{\mu}&\sum_{\xi,\sigma=\pm}\sum_{k=1}^\infty\frac{1}{(\pi k)^2}{\rm cos}\left(\frac{\pi k(\mu^2-\Delta_{\xi\sigma}^2)}{E_1^2}\right)\notag\\
-\frac{e}{h}\sum_{\xi,\sigma=\pm}\frac{\mu^2-\Delta_{\xi\sigma}^2}{2\mu}&\sum_{k=1}^\infty\frac{1}{\pi k}{\rm sin}\left(\frac{\pi k(\mu^2-\Delta_{\xi\sigma}^2)}{E_1^2}\right),
\end{align}
where 
\begin{align}\label{C-Delta}
\mathcal{C}(\Delta_{\xi\sigma},B)&=\frac{|\Delta_{\xi\sigma}|}{2}+\frac{3E_1}{\sqrt{2}}\zeta\left(-\frac{1}{2}, 1+\frac{\Delta_{\xi\sigma}^2}{2E_1^2}\right)\notag\\
&-\frac{\Delta_{\xi\sigma}^2}{2\sqrt{2}E_1}\zeta\left(\frac{1}{2}, 1+\frac{\Delta_{\xi\sigma}^2}{2E_1^2}\right),
\end{align}
and we have used the relation
\begin{align}
\frac{d}{dx}\zeta(s,x)=-s\zeta(s+1,x).
\end{align}
Comparing Eqns.~\eqref{Mag-reg}, and \eqref{Mag-vac}, it is clear that for $\mu>|\Delta_{\xi\sigma}|$, the regular, and vacuum contributions to the magnetization cancel with each other, and only the oscillating piece remains.  For small $B$ fields, we use the expansion\cite{Sharapov:2004}
\begin{align}\label{zeta-approx}
\zeta(s,1+a)\approx \frac{a^{1-s}}{s-1}-\frac{1}{2a^s}+\frac{s}{12a^{1+s}}
\end{align}
for large $a$ to obtain
\begin{align}\label{Mag-approx}
M_{\rm vac}(\mu=0)\approx -\sum_{\xi,\sigma=\pm}\frac{e}{h}\mathcal{M}\left(\Delta_{\xi\sigma},B\right),
\end{align}
where
\begin{align}
\mathcal{M}\left(\Delta_{\xi\sigma},B\right)=\left\lbrace\begin{array}{cc}
\displaystyle\frac{E_1^2}{6|\Delta_{\xi\sigma}|}, & \Delta_{\xi\sigma}\neq 0\\
 & \\
\displaystyle\frac{3E_1}{4\sqrt{2}\pi}\zeta\left(\frac{3}{2}\right), & \Delta_{\xi\sigma}=0
\end{array}\right.,
\end{align}
and $\zeta(x)$ is the Riemann zeta function.  Again, the regular part is the same as the vacuum contribution up to a factor of $-\Theta(\mu-|\Delta_{\xi\sigma}|)$. Plots of the vacuum magnetization as a function of $B$, and $\Delta_z$ are given in Figs.~\ref{fig:Mag-Vac}(a), and (b), respectively.  
For $\Delta_{\rm so}=\Delta_z=0$, we recover the graphene limit.  In this case, the full solution for $M_{\rm vac}$, given by Eqn.~\eqref{Mag-vac}, and shown as a solid black curve in Fig.~\ref{fig:Mag-Vac}(a), agrees perfectly with the form given in Eqn.~\eqref{Mag-approx} (dotted yellow) as it should since no approximations were made when deriving Eqn.~\eqref{Mag-approx} from Eqn.~\eqref{Mag-vac} for $\Delta_{\xi\sigma}=0$.  As a function of $B$, $M_{\rm vac}\propto-\sqrt{B}$ via the definition of $E_1$.  This dependence is evident in the figure.  Since the VSPM case has a single graphene-like band, and one massive band at each valley, for low $B$, there is mainly a $-\sqrt{B}$ dependence, and good agreement with Eqn.~\eqref{Mag-approx} over the range of $B$ shown.  This arises from the fact that the energy scale associated with these small values of $B$ is largely sampling the graphene band, and not the massive band.  So, as $B\rightarrow 0$, the dashed red curve [Eqn.~\eqref{Mag-approx}] is in good agreement with the full result of Eqn.~\eqref{Mag-vac}, and is reduced by approximately a factor of two reflecting that there is only one spin-polarized linear band at each $K$-point.  With increasing field, the massive bands begin to contribute, and Eqn.~\eqref{Mag-approx} begins to deviate from the full result.  The energy scale for such deviations will be controlled by $\Delta_{\xi\sigma}^2\approx 2E_1^2$.  Finally, consider the Kane-Mele QSHI given by $\Delta_z=0$ but finite $\Delta_{\rm so}$ [solid green curve in Fig.~\ref{fig:Mag-Vac}(a)].  In this case, the result of Eqn.~\eqref{Mag-approx} depends on the approximation $a\equiv\Delta_{\xi\sigma}^2/(2E_1^2)=\Delta_{\rm so}^2/(2E_1^2)\gg 1$, and predicts $M_{\rm vac}\propto-B$.  This will only be a good approximation for small $B\ll\Delta_{\rm so}^2/(2\hbar ev^2)$ as seen in Fig.~\ref{fig:Mag-Vac}(a) by comparing the solid green (exact), and dash-double-dotted purple [approximate Eqn.~\eqref{Mag-approx}] curves.  Thus, we expect that as $B\rightarrow 0$, the simple formulae will be robust, and useful for calculating the magnetic susceptibility ($\partial M/\partial B|_{B\rightarrow 0}$).  If one is at large $B$, the full expression must be applied.

Continuing to Fig.~\ref{fig:Mag-Vac}(b), the vacuum magnetization is shown as a function of $\Delta_z$ for three values of magnetic field.  Recall that the TI regime is for $\Delta_z<\Delta_{\rm so}=8$ meV, and the BI case corresponds to $\Delta_z>\Delta_{\rm so}$.  Hence, the divergent features seen at $\Delta_z=\Delta_{\rm so}$ are occurring as the VSPM phase is approached.  In the full calculation for the vacuum contribution (solid curves), no divergence occurs but rather there is a cusp in the magnetization that indicates the critical point between the TI, and BI; this marks the VSPM.  As already understood from the discussion surrounding Fig.~\ref{fig:Mag-Vac}(a), for very low $B$, and small $\Delta_z$, the approximate Eqn.~\eqref{Mag-approx} works reasonably well even as $\Delta_z$ is varied away from zero.  As $\Delta_z\rightarrow\Delta_{\rm so}$, however, the $1/|\Delta_{\xi\sigma}|$ factor in Eqn.~\eqref{Mag-approx} diverges.  This is due to a breakdown in the approximation $\Delta_{\xi\sigma}^2\gg 2E_1^2$ that was used in Eqn.~\eqref{zeta-approx}.  While Eqn.~\eqref{Mag-approx} works well only for very small magnetic fields, and $\Delta_z$ away from the VSPM critical point, it also is seen to work well for very large $\Delta_z$ where $\Delta_{\xi\sigma}^2\gg 2E_1^2$ is again satisfied.  Finally, in order to reconcile the good agreement between Eqn.~\eqref{Mag-approx}, and the VSPM curve in Fig.~\ref{fig:Mag-Vac}(a) with the divergence in Fig.~\ref{fig:Mag-Vac}(b), note that the data point $\Delta_z=\Delta_{\rm so}$ in Fig.~\ref{fig:Mag-Vac}(b) evaluated by Eqn.~\eqref{Mag-approx} is not shown; it would be a discontinuous point that sits near the cusp position in good agreement with the full solution.   In summary, we caution against using the simple formula of Eqn.~\eqref{Mag-approx} beyond its intended region of validity determined by $(\Delta_{\rm so}-\Delta_z)^2\gg 2\hbar ev^2 B $.   We also note that the VSPM state is identified by a cusp (not a divergence) in the vacuum magnetization as the electric field $E_z$ is varied.

\begin{figure}[h!]
\begin{center}
\includegraphics[width=1.0\linewidth]{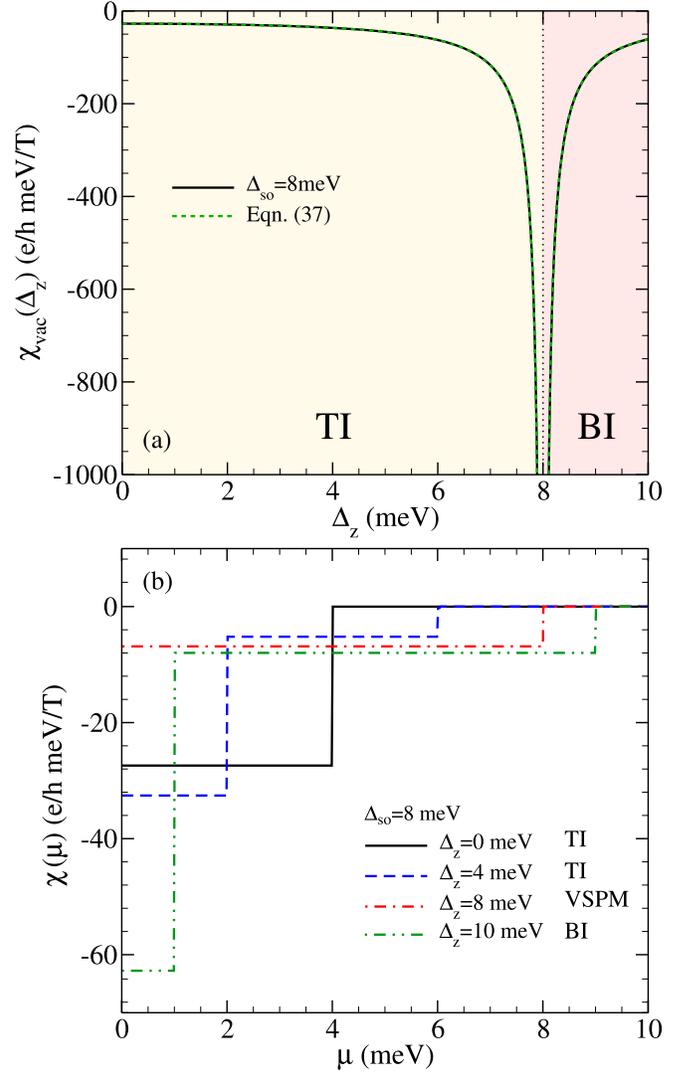}
\end{center}
\caption{\label{fig:Susc}(Color online) (a) Vacuum contribution to the magnetic susceptibility as a function of $\Delta_z$.  Agreement is found between the full result (solid black), and Eqn.~\eqref{Susc-vac} (dashed green). Unlike the magnetization, there is a singularity at $\Delta_z=\Delta_{\rm so}$.  (b) Vacuum plus regular parts of the susceptibility as a function of $\mu$ for various $\Delta_z$.  Steps are seen at the gap values $\Delta_{\rm min}$, and $\Delta_{\rm max}$.
}
\end{figure}

One can also calculate the magnetic susceptibility ($-\partial^2\Omega/\partial B^2|_{B\rightarrow 0}=\chi$).  Differentiating Eqn.~\eqref{Mag-approx} with respect to $B$, and taking the limit $B\rightarrow 0$, we obtain the vacuum susceptibility
\begin{align}\label{Susc-vac}
\chi_{\rm vac}=-\frac{e}{h}\sum_{\xi,\sigma=\pm}\mathcal{X}(\Delta_{\xi\sigma}),
\end{align}
where
\begin{align}
\mathcal{X}(\Delta_{\xi\sigma})=\left\lbrace\begin{array}{cc}
\displaystyle\frac{\hbar ev^2}{6|\Delta_{\xi\sigma}|}, & \Delta_{\xi\sigma}\neq 0\\
 & \\
\displaystyle\lim_{B\rightarrow 0}\frac{3}{8\sqrt{2}\pi}\sqrt{\frac{\hbar ev^2}{B}}\zeta\left(\frac{3}{2}\right), & \Delta_{\xi\sigma}=0
\end{array}\right..
\end{align}
For $\Delta_{\xi\sigma}=0$, Eqn.~\eqref{Susc-vac} diverges as $1/\sqrt{B}$ for $B\rightarrow 0$\cite{McClure:1956,Safran:1979,Principi:2010,Raoux:2014}.  While for finite $\Delta_{\xi\sigma}\rightarrow 0$, $\chi_{\rm vac}$ diverges\cite{Koshino:2011} as $1/|\Delta_{\xi\sigma}|$.  The vacuum plus regular part of the susceptibility is given by
\begin{align}\label{Susc-VR}
\chi=-\frac{e}{h}\sum_{\xi,\sigma=\pm}\mathcal{X}(\Delta_{\xi\sigma})\left[1-\Theta(\mu-|\Delta_{\xi\sigma}|)\right].
\end{align} 
A plot of the vacuum susceptibility as a function of $\Delta_z$ is given in Fig.~\ref{fig:Susc}(a).  The result of differentiating Eqn.~\eqref{Mag-vac} [solid black curve] is compared with Eqn.~\eqref{Susc-vac} [dashed green curve].  
Here, agreement is found between the full result, and that obtained from the approximate equation for $M_{\rm vac}$.  Unlike the magnetization, there is a singularity at $\Delta_z=\Delta_{\rm so}$.  This is due to the massless Dirac fermions of the lowest band.  Since the susceptibility is inversely proportional to the effective mass\cite{McClure:1956,Fukuyama:1971,Safran:1979,Sharapov:2004,Gomez:2011,Principi:2010, Raoux:2014,Koshino:2011}, a divergent diamagnetic response occurs when the VSPM phase is approached\cite{Ezawa:2012d}.  As the two gaps become larger (and the effective mass of the electrons increases), the susceptibility decreases in magnitude.  The slope of the susceptibility has a different sign in the two insulating regimes which may be used to distinguish the two gapped phases.  The contribution to the susceptibility from both the vacuum, and regular parts of $\Omega(\mu)$ as a function of $\mu$ is shown in Fig.~\ref{fig:Susc}(b).  As in gapped graphene\cite{Nakamura:2007,Koshino:2007}, the susceptibility is not zero when $\mu$ is in the gap since the filled valence band provides a constant contribution\cite{Koshino:2007}. For $\Delta_z=0$ (solid black curve), there is only one gap, and the susceptibility is given by its vacuum value until $\mu=\Delta_{\rm so}/2$.  As the chemical potential enters the conduction band, a step occurs, and the susceptibility is zero.  This agrees with the result of doped graphene\cite{Koshino:2007,Koshino:2011}.  For finite $\Delta_z\neq\Delta_{\rm so}$ (dashed blue, and dash-double-dotted green curves), the system is characterized by two gaps.  For $\mu$ less than the minimum gap, the magnitude of the susceptibility is maximized.  Two steps are seen at the values of the gaps $\Delta_{\rm min}$ and $\Delta_{\rm max}$.  Again, for $\mu$ greater than the maximum gap, the susceptibility is zero. In the VSPM phase (dash-dotted red curve), there is only one gap and, thus, a single step.  

The behaviour of the susceptibility should provide a probe of the important energy scales of the system. By varying $\Delta_z$, one can identify the VSPM transitions\cite{Ezawa:2012d} and, thus, $\Delta_{\rm so}$.  In addition, by changing $\mu$, the two gaps $\Delta_{\rm min}$, and $\Delta_{\rm max}$ can be determined.

\subsection{Impurity and Finite Temperature Effects}\label{subsec:Damping-I}

The Landau level broadening that results from impurity scattering ($\Gamma$), and finite temperature can be included by convolving the magnetization with the scattering, and temperature functions
\begin{align}
P_\Gamma(\omega-\mu)=\frac{\Gamma}{\pi[(\omega-\mu)^2+\Gamma^2]},
\end{align}
and
\begin{align}
P_T(\omega-\mu)=-\frac{\partial n_F}{\partial\omega}=\frac{1}{4T{\rm cosh}^2\left(\frac{\omega-\mu}{2T}\right)},
\end{align}
respectively, where $n_F$ is the Fermi distribution function.  For simplicity, we have assumed (as done in Ref.~\cite{Sharapov:2004} and other works) that $\Gamma$ is the same for all Landau levels.  More complicated
descriptions of impurity scattering exist; in particular, see Ref.~\cite{Koshino:2007} which deals explicitly with the diamagnetism of disordered graphene.  Therefore,
\begin{align}\label{Mag-Imp-T}
M(\mu,\Gamma, T)=\int_{-\infty}^\infty\int_{-\infty}^\infty P_\Gamma(\omega-\mu)P_T(\omega^\prime-\mu)M(\omega^\prime)d\omega^\prime d\omega,
\end{align}
where $M(\omega)$ is obtained by setting $\mu=\omega$ in the $\Gamma=0$, and $T=0$ result [Eqns.~\eqref{Mag-reg}-\eqref{Mag-osc}].

Let us begin by examining the effect of impurity scattering on the vacuum plus regular piece of the magnetization.  Combining Eqns.~\eqref{Mag-reg}, and ~\eqref{Mag-vac}, Eqn.~\eqref{Mag-Imp-T} gives
\begin{align}
M(\mu,\Gamma)&=\frac{e}{h}\sum_{\xi,\sigma=\pm}\mathcal{C}(\Delta_{\xi\sigma},B)\int_{-|\Delta_{\xi\sigma}|}^{|\Delta_{\xi\sigma}|}\frac{\Gamma}{\pi[(\omega-\mu)^2+\Gamma^2]}d\omega\notag\\
&=\frac{e}{h}\sum_{\xi,\sigma=\pm}\mathcal{C}(\Delta_{\xi\sigma},B)\mathcal{D}_{\Gamma},
\end{align}
where
\begin{align}\label{D-Gamma}
\mathcal{D}_\Gamma=\frac{1}{\pi}\left[{\rm arctan}\left(\frac{2|\Delta_{\xi\sigma}|\Gamma}{\mu^2-\Delta_{\xi\sigma}^2+\Gamma^2}\right)\right],
\end{align}
and we have taken $T=0$.  For our low $B$ expression [see Eqn.~\eqref{Mag-approx}], $\mathcal{C}(\Delta_{\xi\sigma},B)$ should be replaced with $\mathcal{M}(\Delta_{\xi\sigma},B)$.  Similarly, the effect of impurities on the vacuum plus regular part of the susceptibility is [see Eqn.~\eqref{Susc-VR}]
\begin{align}
\chi(\Gamma)=-\frac{e}{h}\sum_{\xi,\sigma=\pm}\mathcal{X}(\Delta_{\xi\sigma})\mathcal{D}_\Gamma.
\end{align}
This agrees with  Eqn.~(6) of Ref.~\cite{Nakamura:2007} in the gapped graphene-limit $|\Delta_{\xi\sigma}|\rightarrow\Delta$.  For both the magnetization, and susceptibility, the damping factor $\mathcal{D}_\Gamma$ has a dependence on chemical potential.  As $\mu$ increases, the magnetic response decreases.

The same procedure is applied when considering the effect of finite temperature.  Equation~\eqref{Mag-Imp-T}, with the combined result of Eqns.~\eqref{Mag-reg}, and \eqref{Mag-vac}, gives
\begin{align}
M(\mu,T)&=\frac{e}{h}\sum_{\xi,\sigma=\pm}\mathcal{C}(\Delta_{\xi\sigma},B)\int_{-|\Delta_{\xi\sigma}|}^{|\Delta_{\xi\sigma}|}\left(-\frac{\partial n_F}{\partial\omega}\right)d\omega\notag\\
&=\frac{e}{h}\sum_{\xi,\sigma=\pm}\mathcal{C}(\Delta_{\xi\sigma},B)\mathcal{D}_T,
\end{align}
where
\begin{align}\label{DT}
\mathcal{D}_T=\frac{{\rm sinh}(\beta|\Delta_{\xi\sigma}|)}{{\rm cosh}(\beta|\Delta_{\xi\sigma}|)+{\rm cosh}(\beta\mu)},
\end{align}
$\beta=T^{-1}$, and $\Gamma$ has been set to zero.  While we have provided the general result here, there are limiting cases which are known, and discussed below.   For $\Delta_{\xi\sigma}\rightarrow 0$, sinh$(\beta|\Delta_{\xi\sigma}|)\approx\beta|\Delta_{\xi\sigma}|$ and
\begin{align}\label{DT-D0}
\mathcal{D}_T\approx\frac{\beta|\Delta_{\xi\sigma}|}{1+{\rm cosh}(\beta\mu)}.
\end{align}
The magnetic susceptibility is
\begin{align}
\chi(T)=-\frac{e}{h}\sum_{\xi,\sigma=\pm}\mathcal{X}(\Delta_{\xi\sigma})\mathcal{D}_T.
\end{align}
Using Eqn.~\eqref{DT-D0}, the susceptibility for $\Delta_{\xi\sigma}\rightarrow 0$ is
\begin{align}
\chi(T)=-\frac{e}{h}\sum_{\xi,\sigma=\pm}\frac{\hbar ev^2}{12}\frac{\beta}{{\rm cosh}^2\left(\beta\mu/2\right)},
\end{align}
which is in agreement with the results of Refs.~\cite{McClure:1956,Safran:1979,Principi:2010}.  Returning to Eqn.~\eqref{DT}, for zero chemical potential,
\begin{align}
D_T=\rm{tanh}\left(\beta|\Delta_{\xi\sigma}|\right),
\end{align}
and
\begin{align}
\chi(T)=-\frac{e}{h}\sum_{\xi,\sigma=\pm}\mathcal{X}(\Delta_{\xi\sigma})\rm{tanh}\left(\beta|\Delta_{\xi\sigma}|\right).
\end{align}
This is in agreement with Eqn.~(31) in Ref.~\cite{Ezawa:2012d}.

\begin{figure}[h!]
\begin{center}
\includegraphics[width=1.0\linewidth]{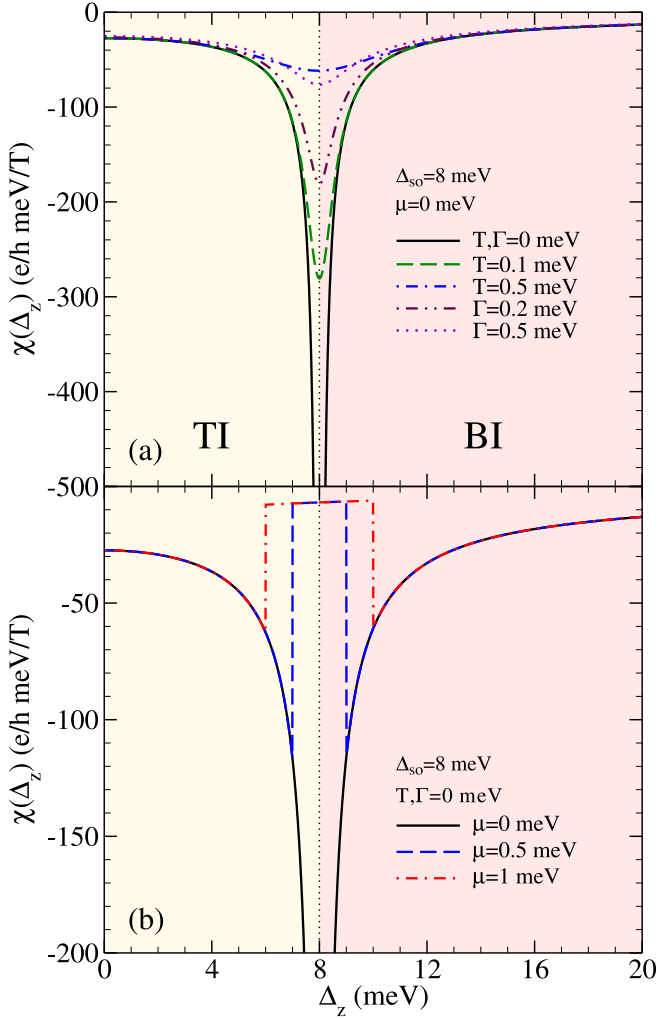}
\end{center}
\caption{\label{fig:Susc-T}(Color online) (a) The effect of finite temperature, and impurity scattering on the susceptibility as a function of $\Delta_z$. (b) The effect of finite chemical potential on the susceptibility.
}
\end{figure}

The $\mu=0$ susceptibility at finite temperature, and impurity scattering is shown in Fig.~\ref{fig:Susc-T}(a) as a function of $\Delta_z$.
At $T=\Gamma=0$ (solid black curve), a singularity exists at the critical value $\Delta_{\rm so}=\Delta_z$.  The singularity is removed by the inclusion of finite temperature or impurity effects\cite{Principi:2010}.  Note that the two curves for either temperature (dash-dotted blue) or impurity scattering (dotted magenta) equal to 0.5 meV are very similar.  While the singularity is washed out, there is a clear minimum at the position of the phase transition.  This is independent of magnetic field since we are dealing with $\chi$.  Of course, the impurity scattering rate itself depends on the type of impurities\cite{Neto:2009}. Returning to Eqn.~\eqref{Susc-VR}, we note that finite $\mu$ will also remove the singularity.  This is shown in Fig.~\ref{fig:Susc-T}(b).  There is an abrupt vertical drop in the absolute value of $\chi$ as the magnitude of $\Delta_{\xi\sigma}$ is reduced below the value of $\mu$.  For $\Delta_{\rm max}>\mu>\Delta_{\rm min}$, the susceptibility remains finite, and is given by $\chi=-[e/h][\hbar ev^2/(3\Delta_{\rm max})]$.  This value originates entirely from the spin-valley branches of the Dirac dispersion which have the maximum gap $\Delta_{\rm max}$.  This abrupt change in $\chi$ can be used to determine the size of $\mu$ (i.e. the doping away from charge neutrality).  The determining structures involved in such a measurement are most prominent for small values of chemical potential.

\section{Magnetic Oscillations}

\subsection{$T=\Gamma=0$}

\begin{figure}[]
\begin{center}
\includegraphics[width=0.90\linewidth]{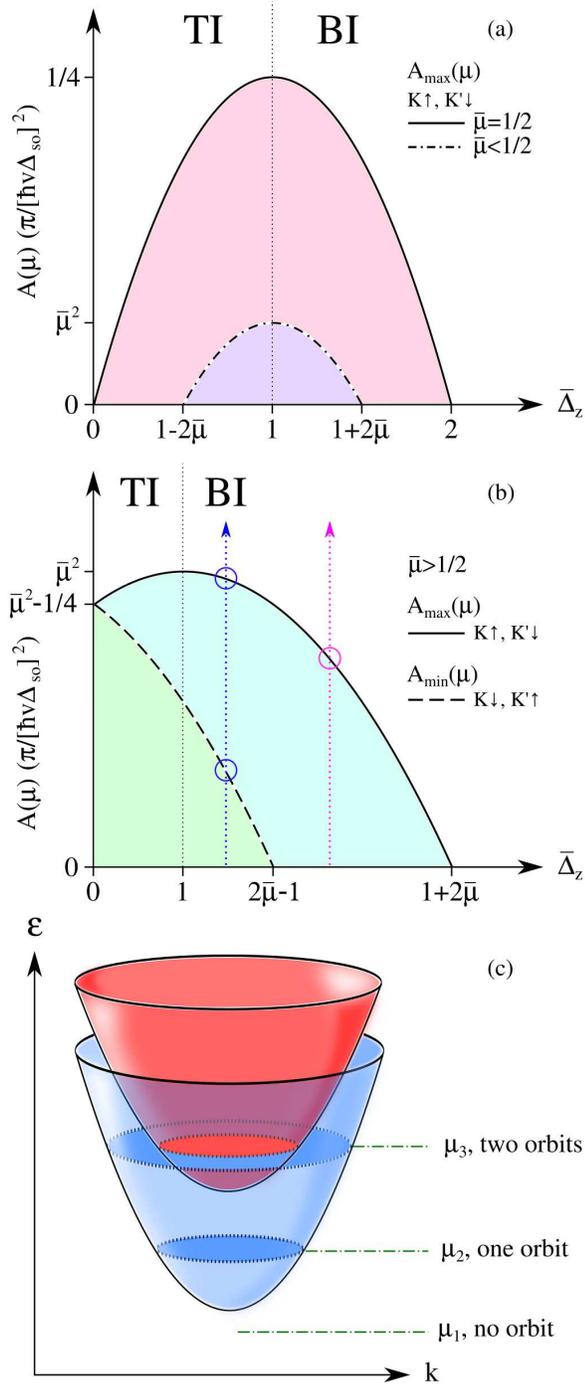}
\end{center}
\caption{\label{fig:Area}(Color online) (a)  The cyclotron orbit area as a function of $\mu$ for $\mu$ between the two gaps.  In this regime, a single cyclotron orbit exists for $1-2\bar{\mu}<\bar{\Delta}_z<1+2\bar{\mu}$. Note: $\bar{\mu}\equiv\mu/\Delta_{\rm so}$, and $\bar{\Delta}_z\equiv\Delta_z/\Delta_{\rm so}$.  (b) The two cyclotron orbit areas for $\mu$ above both gaps.     (c) Schematic representation of the positive energy band structure at the $K$ point.  For $\mu=\mu_1$, no cyclotron orbits are present.  For $\mu=\mu_2$ [see (a)], only one orbit exists.  For $\mu=\mu_3$ [see (b)], two orbits exist. 
}
\end{figure}

\begin{figure}[h!]
\begin{center}
\includegraphics[width=1.0\linewidth]{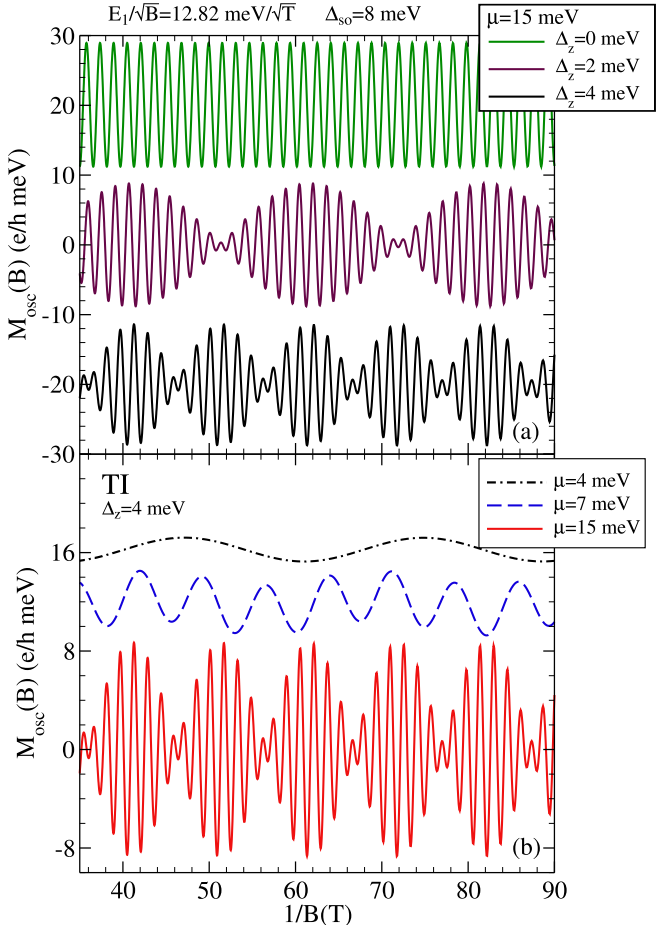}
\end{center}
\caption{\label{fig:Mosc}(Color online) Magnetic oscillations as a function of $1/B$. (a) dHvA effect for $\mu=15$ meV (above all the gaps).  For finite $\Delta_z$, two orbit areas exist, and a beating phenomenon emerges. (b)  $M_{\rm osc}(B)$ for $\Delta_z=4$ meV, and varying $\mu$.  For clarity, the curves have been offset from oscillating around zero. 
}
\end{figure}

We now return to Eqn.~\eqref{Mag-osc} to discuss the magnetic oscillations in a doubly gapped Dirac system like silicene.  The de-Haas van-Alphen (dHvA) effect is a low-field phenomenon which provides an experimental probe of the Fermi surface.  Semiclassically, $M_{\rm osc}$ is found to oscillate according to\cite{Luk:2004, Suprunenko:2008, Wright:2013}
\begin{align}
M^k_{\rm osc}\propto{\rm sin}\left(2\pi k\left[\frac{\hbar A(\mu)}{2\pi eB}-\gamma\right]\right),
\end{align}
where $A(\mu)$ is the area of a cyclotron orbit at the Fermi energy ($A(\mu)=\pi k_F^2$), and $\gamma$ is a phase offset related to the Berry's phase ($\gamma=1/2$ for a two-dimensional electron gas, and $0$ for a relativistic Dirac-like system with Berry's phase $\pi$).  For silicene, $\gamma=0$, and there are two possible orbit areas.  In silicene, the Fermi momentum for the band labelled by $\xi$ and $\sigma$ is
\begin{align}
k_F^{\xi\sigma}=\sqrt{\frac{\mu^2-\Delta_{\xi\sigma}^2}{\hbar^2v^2}},
\end{align}
where only real solutions are considered.  Therefore,
\begin{align}\label{Area}
A_{\xi\sigma}(\mu)=\frac{\pi}{\hbar^2v^2}\left[\mu^2-\frac{\left(\Delta_{\rm so}-\xi\sigma\Delta_z\right)^2}{4}\right].
\end{align}
Indeed, taking only the leading term of Eqn.~\eqref{Mag-osc}, we note
\begin{align}\label{Mosc-area}
M_{\rm osc}(\mu)=-\frac{e}{h}\sum_{\xi,\sigma=\pm}\frac{\mu^2-\Delta_{\xi\sigma}^2}{2\mu}\sum_{k=1}^\infty\frac{1}{\pi k}{\rm sin}\left(2\pi k\left[\frac{\hbar A_{\xi\sigma}(\mu)}{2\pi eB}-\gamma\right]\right),
\end{align}
where $\gamma=0$, and $A_{\xi\sigma}(\mu)$ is given by Eqn.~\eqref{Area}.  Equation~\eqref{Area} is plotted in Fig.~\ref{fig:Area}(a) for $\bar{\mu}\equiv\mu/\Delta_{\rm so}\leq 1/2$ as a function of $\bar{\Delta}_z\equiv\Delta_z/\Delta_{\rm so}$.
The results are shown for the lowest gapped bands (spin-up at $K$, and spin-down at $K^\prime$) since, for $\bar{\mu}\leq 1/2$, only one band is occupied.  For $\bar{\mu}<1/2$ (dash-dotted curve), the lowest band is occupied (and there is an associated cyclotron orbit) if $1-2\bar{\mu}<\bar{\Delta}_z<1+2\bar{\mu}$.  As $\mu$ is increased, the range of $\Delta_z$ grows for which $A(\mu)$ is nonzero.  The maximum orbit area is fixed at $\bar{\mu}^2$, and occurs at $\Delta_z=\Delta_{\rm so}$ (the VSPM phase).  $\bar{\mu}=1/2$ (solid curve) corresponds to the maximum range of $\Delta_z$ for which $A(\mu)>0$ when only one band is occupied.  The lowest band remains occupied between $\bar{\Delta}_z=0$, and 2.  In all cases, there is perfect symmetry between the TI, and BI regimes.  Schematically, the regions of $\Delta_z$ which give a finite area are those which ensure $\mu$ sits in the $\mu_2$ position of Fig.~\ref{fig:Area}(c).  $\mu=\mu_1$ corresponds to no cyclotron orbits.  Next, consider $\bar{\mu}>1/2$.  This corresponds to $\mu=\mu_3$ in Fig.~\ref{fig:Area}(c).  Now both bands are occupied.  $A(\mu)$ as a function of $\bar{\Delta}_z$ is shown in Fig.~\ref{fig:Area}(b).  The maximum area (solid curve) comes from the spin-up electrons at $K$, and the spin-down electrons at $K^\prime$.  The other spin-valley combinations make up the minimum area (dashed curve).   For $A_{\rm max}(\mu)$, the maximum value of $\bar{\Delta}_z$ remains at $1+2\bar{\mu}$.  $1-2\bar{\mu}$ now occurs for negative $\bar{\Delta}_z$; thus, for $\Delta_z=0$, $A_{\rm max}(\mu)$ is finite at a value of $\bar{\mu}^2-1/4$.  The maximum still occurs in the VSPM phase, and retains its value of $\bar{\mu}^2$.  $A_{\rm min}(\mu)$ persists until $\bar{\Delta}_z=2\bar{\mu}-1$, and is equal to $A_{\rm max}(\mu)$ at $\Delta_z=0$.  There is perfect symmetry between the maximum, and minimum areas if negative values of $\bar{\Delta}_z$ are considered.  In that region, the spin, and valley labels of the two areas are interchanged.  It is clear that for $\bar{\Delta}_z<2\bar{\mu}-1$, there are two orbits which will contribute to the magnetic oscillations.  For $\Delta_z=0$, the areas are equal, and will contribute an overall degeneracy to $M_{\rm osc}$. For finite $\Delta_z$, the difference in the two orbits increases with the electric field strength.  The interaction between the two areas is clearly seen by looking at the total oscillating magnetization [Eqn.~\eqref{Mosc-area}].  We are only interested in the lowest harmonic $k=1$.  Therefore,
 \begin{align}\label{Mosc-full}
M_{\rm osc}(\mu)=-\frac{e}{h}\left\lbrace\frac{\mu^2-\Delta_{\rm min}^2}{\pi\mu}{\rm sin}\left(2\pi\left[\frac{\hbar A_{\rm min}(\mu)}{2\pi eB}\right]\right)\right.\notag\\
\left.+\frac{\mu^2-\Delta_{\rm max}^2}{\pi\mu}{\rm sin}\left(2\pi\left[\frac{\hbar A_{\rm max}(\mu)}{2\pi eB}\right]\right)\right\rbrace,
\end{align}
where $A_{\rm min}\equiv A_{+-}$, and $A_{\rm max}\equiv A_{++}$.  Equation~\eqref{Mosc-full} shows that $M_{\rm osc}$ is constructed from two sine waves with $1/B$ frequencies of
\begin{align}
\omega_{\rm min}=\frac{\hbar A_{\rm min}(\mu)}{e}=\frac{\pi}{\hbar ev^2}\left[\mu^2-\frac{\left(\Delta_{\rm so}+\Delta_z\right)^2}{4}\right],
\end{align}
and
\begin{align}
\omega_{\rm max}=\frac{\hbar A_{\rm max}(\mu)}{e}=\frac{\pi}{\hbar ev^2}\left[\mu^2-\frac{\left(\Delta_{\rm so}-\Delta_z\right)^2}{4}\right].
\end{align}
Therefore, for $\mu$ between the two gaps [$\mu_2$ in Fig.~\ref{fig:Area}(c)], there is only one characteristic dHvA frequency.  For $\mu$ above both gaps [$\mu_3$ in Fig.~\ref{fig:Area}(c)], there are two frequencies, and interference will be observed\cite{Islam:2014,Mawrie:2014,Tsaran:2014}.  When the difference between $A_{\rm min}(\mu)$, and $A_{\rm max}(\mu)$ is small, a strong beating is present.  Looking at Fig.~\ref{fig:Area}(b), this will occur when $\Delta_z$ is very small (but nonzero) or $\mu$ is very large ($\mu\gg\Delta_{\rm max}$). The beating of the quantum oscillations is shown in Fig.~\ref{fig:Mosc}.
Figure~\ref{fig:Mosc}(a) shows $M_{\rm osc}$ as a function of $1/B$ for $\mu=15$ meV, and $\Delta_z=0$, 2, and 4 meV (green, purple, and black curves, respectively).  Here $\mu$ is above both gaps.  For $\Delta_z=0$ meV, only one gap is present, and strong dHvA oscillations are observed.  For $\Delta_z=2$ meV, the bands have become spin split, and a beating phenomenon emerges\cite{Islam:2014,Tsaran:2014} in addition to the usual oscillations.  For $\Delta_z=4$ meV, the beating persists; however, the beat frequency has increased.  This phenomenon is a signature of a small difference between the two orbit areas which we control here with the magnitude of the external electric field.  For smaller values of $\mu$, increasing $\Delta_z$ can lead from a single fundamental frequency to a beating regime, and back to a single frequency as $\Delta_z$ becomes large enough that $\mu$ is only above the smaller gap.

The magnetic oscillations for fixed $\Delta_z=4$ meV, and varying $\mu$ are shown in Fig.~\ref{fig:Mosc}(b).  For $\mu=4$ meV (dash-dotted black curve) only a single frequency is present.  When $\mu=7$ meV (dashed blue curve), both bands contribute, and two characteristic frequencies become visible.  They are $\omega_{\rm max}=0.860$T, and $\omega_{\rm min}=0.249$T.  As $\mu$ is increased further (solid red curve), the two cyclotron orbits become closer ($\omega_{\rm max}=4.22$T, and $\omega_{\rm min}=3.61$T), and beating is clearly observed.  Analogous results are found for the VSPM, and BI; however, for the VSPM, dHvA oscillations are seen for all $\mu\neq 0$.

\subsection{Impurity and Finite Temperature Effects}

In the presence of impurities, the magnetic oscillations are damped by a Dingle factor which depends on the scattering probability.  This effect is included by convolving $M_{\rm osc}(\mu)$ with $P_\Gamma(\mu-\Gamma)$ as is shown in Sec.~\ref{subsec:Damping-I}.  For gapped graphene, this has been worked out analytically\cite{Sharapov:2004}, and can be extended to the doubly gapped case of silicene.  Thus, in the presence of impurities, and at finite temperature [see Eqn.~(8.17) of Ref.~\cite{Sharapov:2004} which we repeat here for completeness],
\begin{align}
\Omega_{\rm osc}^{\xi\sigma}(\mu,\Gamma, T)&=\frac{eBE_1^2}{2h\mu}\sum_{k=1}^\infty\frac{1}{(\pi k)^2}{\rm cos}\left(\frac{\pi k(\mu^2-\Delta_{\xi\sigma}^2-\Gamma^2)}{E_1^2}\right)\notag\\
&\times\mathcal{R}_D\mathcal{R}_T,
\end{align}
with the Dingle factor
\begin{align}
\mathcal{R}_D={\rm exp}\left(-2\pi k\frac{\mu\Gamma}{E_1^2}\right),
\end{align}
and temperature factor
\begin{align}
\mathcal{R}_T=\frac{k\lambda}{{\rm sinh}(k\lambda)},
\end{align}
where
\begin{align}
\lambda=\frac{2\pi^2T\mu}{E_1^2}.
\end{align}
For both impurity, and temperature effects, the damping of the oscillations is dependent on $\mu$ making the dHvA effect difficult to observe for large values of chemical potential\cite{Sharapov:2004}.

\section{Conclusions}

The application of a perpendicular electric field to a low-buckled honeycomb lattice creates an onsite potential difference ($\Delta_z$) between the two sublattices.  The resulting electronic band structure consists of four valley-spin polarized bands which result from a large spin-orbit interaction.  An example material is silicene (a monolayer of silicon atoms).  The Dirac fermions involved are massive with the magnitude of the two resulting gaps modulated by $\Delta_z$.  By varying the electric field, a phase transition from a topological insulator, through a valley-spin polarized metal, to a trivial band insulator is induced.  We have studied the magnetization of such bands in an attempt to find distinctive features associated with the three different phases.  Particular attention is given to identifying the boundary between the TI, and BI.

The derivative of the magnetization ($M$) with respect to the chemical potential ($\mu$) is calculated, and found to give a series of steps as a function of $\mu$.  The onset of each new plateau is associated with the energies of the Landau levels involved.  Through direct calculation of the DC Hall conductivity ($\sigma_H$), we have verified explicitly that the Streda formula $e(\partial M/\partial\mu)=\sigma_H$ holds.  In the absence of Zeeman splitting, the Hall conductivity (in units of $e^2/h$) has filling factors $\nu=0,\pm 1, \pm 2, \pm 4, \pm 6,...$ for both the TI, and BI; while, for the VSPM, $\nu=\pm 1, \pm 2, \pm 4, \pm 6,...$.  Including a finite Zeeman interaction splits the valley-spin degeneracy of the $\nu=\pm 4, \pm 6,...$ filling factors, and the resulting Hall conductivity can take any integer multiple of $e^2/h$.  For moderate values of Zeeman splitting, the four steps associated with each of the $N=1,2,3,...$ levels become spin-polarized.  Therefore, as an example, while the $\nu=2$ Hall conductivity is spin-degenerate, the $\nu=3,4$ conductivities have a net spin-up imbalance.  The spin polarization is lost for $\nu=5,6$, and then regained for $\nu=7,8$, etcetera.  Near the charge neutral point, $\sigma_H$ can also display spin, and valley polarization even when the Zeeman splitting is neglected.  We emphasize that the boundary between the TI, and BI supports a finite charge Hall conductivity providing a signature of the VSPM.  This conductivity goes from $-e^2/h$ to $e^2/h$ at $\mu=0$ while it is zero for the other two phases.  The finite charge Hall conductivity near $\mu=0$ of the VSPM is lost as the Zeeman interaction is increased.

Steps analogous to those found in the derivative of the magnetization also exist in the integrated density of states [$I(\omega_{\rm max})$] to energy $\omega_{\rm max}$, as well as in the optical spectral weight [$W(\Omega_c)$] accumulated below a cutoff frequency $\Omega_c$.  In the first instance, the energy at which the new steps onset is determined by the energies of the various Landau levels.  In the optical case, it is the energy of the allowed optical transitions which matters; this involves the sum of two Landau level energies for the interband transitions, and the difference for the intraband transitions.  In both cases, we present simple analytic algebraic expressions for the height of the various plateaus; these are not integral in fundamental constants but involve only the Landau level energies, and the band gaps.  For the quasiparticle density of states in the TI, and BI regimes, $I(\omega_{\rm max})$ is zero for $\omega_{\rm max}$ less then the two gaps.  It is finite in the VSPM for all $\omega_{\rm max}\neq 0$.  This provides a clear signature of the phase transition between the TI, and BI phases.  For $W(\Omega_c)$, conservation of spectral weight is noted for all $\mu$.  While the magnitude of the chemical potential strongly affects the height of the first plateau, for $\Omega_c$ much larger than the gaps, and chemical potential, all curves merge into the $B=0$ background; this has a linear dependence on $\Omega_c$.

The magnetic susceptibility ($\chi$) as a function of the onsite potential difference is found to exhibit a singularity as $\Delta_z$ goes through the spin-orbit band gap $\Delta_{\rm so}$.  This results from the lowest band gap closing as the system enters the VSPM phase, and the well understood divergent graphene result emerges.  The singularity is not present for any other $\Delta_z$, and can be removed by finite temperature or by impurity scattering ($\Gamma$).  For parameters characteristic of silicene, we find that to observe a residual signature of the divergence, it is necessary to have $\Gamma\leq 0.5$ meV or temperatures of a few Kelvin at most.   A finite value of magnetic field also modifies the dependence of the magnetization as a function of $\Delta_z$ providing only a kink or change in slope at $\Delta_z=\Delta_{\rm so}$.  This could still be employed to identify the boundary between the nontrivial, and trivial topological phase.  In the presence of doping away from charge neutrality, the finite chemical potential provides a cutoff on the contribution to the susceptibility that comes from the dispersion branch with the minimum gap $\Delta_{\rm min}$.  It is entirely eliminated for all values of $\Delta_z$ such that $\mu>(1/2)|\Delta_z-\Delta_{\rm so}|$.  This fact can be used to obtain an estimate of the chemical potential as well as the magnitude of the spin-orbit gap $\Delta_{\rm so}$.

Finally, the de-Haas van-Alphen magnetic oscillations are discussed.  Depending on the location of the chemical potential, there can be zero, one, or two fundamental cyclotron frequencies; these occur when $\mu$ is below both gaps, between the gaps, or above both gaps, respectively.  For the VSPM, there is only one gap; thus, there will always be at least one fundamental frequency.  When there are two fundamental cyclotron frequencies, a beating pattern emerges in the oscillations.  For large values of $\mu$, where the Dirac fermion dispersion curves merge, the beating is lost.  For no external potential, the system is a topological insulator, and there is a single degenerate cyclotron frequency which can be split by the application of a small $\Delta_z$.

In summary, we have examined the magnetic response of Dirac fermions in a low-buckled honeycomb lattice.  We have found signatures which should allow for an experimental verification of the two insulating regimes (TI, and BI) and, the VSPM through tuning by an external electric field.

\emph{Note added.}  During the preparation of this manuscript, a preprint appeared\cite{Raoux:2014a} which discusses the effect of finite $\mu$ on the susceptibility of related 2D systems.

\begin{acknowledgments}
This work has been supported by the Natural Sciences and Engineering Research Council of Canada and, in part, by the Canadian Institute for Advanced Research, and by the National Science Foundation under Grant No. NSF PHY11-25915.
\end{acknowledgments}

\bibliographystyle{apsrev4-1}
\bibliography{MO-Si}

\end{document}